\documentclass[prd,twocolumn,showpacs,superscriptaddress,nofootinbib,preprintnumbers,aps]{revtex4}
\usepackage{color}
\usepackage{latexsym, amssymb, amsmath,color,graphicx}
\usepackage[unicode=true, 
 bookmarks=true,bookmarksnumbered=false,bookmarksopen=false,
 breaklinks=false,pdfborder={0 0 1},backref=false,colorlinks=true]
 {hyperref}

 \makeatletter

\@ifundefined{textcolor}{}
{%
 \definecolor{BLACK}{gray}{0}
 \definecolor{WHITE}{gray}{1}
 \definecolor{RED}{rgb}{1,0,0}
 \definecolor{GREEN}{rgb}{0,1,0}
 \definecolor{BLUE}{rgb}{0,0,1}
 \definecolor{CYAN}{cmyk}{1,0,0,0}
 \definecolor{MAGENTA}{cmyk}{0,1,0,0}
 \definecolor{YELLOW}{cmyk}{0,0,1,0}
 }
 \def\bea{\begin{eqnarray}}
 \def\eea{\end{eqnarray}}

\usepackage{hyperref}
\hypersetup{
    colorlinks,%
    citecolor=blue,%
    filecolor=blue,%
    linkcolor=blue,%
    urlcolor=blue
}

\begin{document}

\title{High Energy Scattering in Perturbative Quantum Gravity at Next to Leading Power}

\preprint{YITP-SB-13-25}

\author{Ratindranath Akhoury}
\email{akhoury@umich.edu}
\affiliation{Michigan Center for Theoretical Physics, Randall Laboratory of Physics, University of Michigan, Ann Arbor, MI 48109-1120, USA}

\author{Ryo Saotome}
\email{rsaotome0927@gmail.com}
\affiliation{10 Union Ave, Apt 421, Bala Cynwyd PA 19004, USA}

\author{George Sterman}
\email{george.sterman@stonybrook.edu}
\affiliation{C.N. Yang Institute for Theoretical Physics, and Department of Physics and Astronomy, Stony Brook University, Stony Brook, New York, 11794-3840, USA}

\date{\today}

\begin{abstract}

We consider the relativistic scattering of unequal-mass scalar particles through graviton exchange in the small-angle high-energy regime.  We show the self-consistency of expansion around the eikonal limit and compute the scattering amplitude up to the next-to-leading power correction of the light particle energy, including gravitational effects of the same order.   The first power correction is suppressed by a single power of the ratio of momentum transfer to the energy of the light particle in the rest frame of the heavy particle, independent of the heavy particle mass. We find that only gravitational corrections contribute to the exponentiated phase in impact parameter space in four dimensions.     For large enough heavy-particle mass, the saddle point for the impact parameter is modified compared to the leading order by a multiple of the Schwarzschild radius determined by the mass of the heavy particle, independent of the energy of the light particle.

\end{abstract}


\maketitle


\section{Introduction}

There has been a renewed interest in perturbative quantum gravity in part due to recent developments illuminating a relationship between gravity and gauge theory amplitudes \cite{Bern:2008qj, Bern:2010ue}, because of its relevance in high energy, small angle scattering \cite{tHooft:1987vrq, Amati:1987wq, Giudice:2001ce, Giddings:2007bw, DAppollonio:2010krb, Giddings:2011xs, Naculich:2011ry} and for connections to the dynamics revealed in gravitational radiation \cite{TheLIGOScientific:2016agk}. The long distance regime of this theory is particularly interesting as quantum gravity has a simpler infrared structure than gauge theories. For instance, there is a cancellation of collinear divergences of wide-angle scattering amplitudes in the former theory, but not the latter \cite{Weinberg:1965nx, Akhoury:2011kq, Beneke:2012xa} and the logarithmically divergent soft amplitudes have a ladder like structure.

Since the infrared behavior of perturbative quantum gravity is tractable, it is important to look for observables for which long distance effects are particularly important.  A fundamentally infrared dominated physical quantity of note in perturbative quantum gravity is the eikonal phase that has been calculated for small-angle high-energy scattering processes. An eloquent overview of the eikonal regime for this type of scattering is given in a set of lectures by Giddings \cite{Giddings:2011xs}.  In recent years, related considerations have been applied to more general kinematics and observables, using and developing amplitude techniques 
\cite{Bjerrum-Bohr:2013bxa,
Ciafaloni:2014esa,
White:2014qia,
Bjerrum-Bohr:2016hpa,
Luna:2016idw,
Goldberger:2016iau,
Okui:2017all,
Bjerrum-Bohr:2018xdl,
Collado:2018isu,
Kosower:2018adc,
KoemansCollado:2019ggb,
Bjerrum-Bohr:2019kec,
Cristofoli:2020uzm,
AccettulliHuber:2020oou}.
These concepts have been tested by
exploiting new results in generalized theories with gravity 
\cite{Camanho:2014apa,
Hinterbichler:2017qyt,
KoemansCollado:2019lnh,
Glampedakis:2019dqh,
Kulaxizi:2019tkd,
DiVecchia:2019myk,
DiVecchia:2019kta,
Bern:2020gjj},
and have  been extensively applied and compared to classical gravitational scattering of massive objects, especially black holes 
\cite{Damour:2017zjx,
Cheung:2018wkq,
Guevara:2018wpp,
Cristofoli:2019neg,
Churilova:2019jqx,
Bern:2019crd,
Damour:2019lcq,
Bern:2020buy,
Parra-Martinez:2020dzs,
Kalin:2020mvi,
KoemansCollado:2020sxs,
Antonelli:2020ybz,
Mogull:2020sak,
Bern:2020uwk}.

In this paper, we provide a detailed analysis of the expansion around the eikonal limit directly in  perturbation theory for the gravitational scattering of a very light scalar by a heavy scalar.   We verify the diagrammatic self-consistency of this expansion and re-derive next-to-leading corrections in arbitrary dimensions. Although the exponentiation of the leading eikonal phase is a classic result, and the next-to-leading correction has been studied intensively 
\cite{DAppollonio:2010krb,
White:2014qia,
Luna:2016idw,
Collado:2018isu, 
Kulaxizi:2019tkd},
we are not aware of another explicit, all-order diagrammatic treatment of the next-to-leading correction.\footnote{This paper expands previous work made public by the authors on this subject.}   Our analysis applies to arbitrary dimensions, and agrees with previous fixed-order calculations of this correction \cite{DAppollonio:2010krb}.   

The high-energy, small-angle scattering amplitude has been shown to be largely independent of short distance physics. Giddings, Gross and Maharana \cite{Giddings:2007bw} argued that the scattering of strings may be replaced by graviton contributions alone with short distance physics playing no significant role.  As reviewed in \cite{Giddings:2011xs} the leading eikonal approximation is justified for high energy, small angle scattering of equal-mass particles, because the impact parameter as determined from the saddle point of the loop integrations is larger than the typical gravitational radius $2GE$  by a factor of $s/t$. Thus non-local string effects are actually subdominant to higher loop gravitational processes for a large range of impact parameters. In a related analysis, in \cite{DAppollonio:2010krb} the scattering of massless closed strings from a stack of D-branes was explored in various kinematical regimes including the eikonal. In the region of impact parameter large compared to relevant Schwarzschild radius associated with the center of mass energy, this reference also find that gravity effects dominate.  In a kinematic regime related to the latter study, we will derive the first subleading contributions coming from field theory corrections to the eikonal phase, in arbitrary dimensions.

The eikonal approximation provides the leading contribution, which exponentiates in impact parameter space. Large impact parameters dominate the scattering in this regime. The next-to-eikonal contribution to small-angle high-energy gravitational scattering has to our knowledge not previously been calculated in the manner described below, although there has been much interesting progress on this topic for non-abelian gauge theories \cite{Laenen:2008gt,Dukes:2013gea,
Melville:2016bss,
Moult:2017xpp,
Bahjat-Abbas:2019fqa,
Choi:2019rlz,
White:2019ggo}. In \cite{White:2011yy} the effective vertices at the next-to-eikonal level for gravity are given. Our specific calculations will also make contact with the extensive computations of Ref.\ \cite{BjerrumBohr:2002ks}. A number of our results will involve special limits of the diagrammatic calculations outlined there.   

In this paper, we consider the near forward scattering of unequal-mass scalar particles, and compute  power corrections to the eikonal approximation in ${\Delta \over E_\phi}$, where $\Delta$ is the momentum transfer and $E_\phi$ the energy of a light projectile that undergoes small-angle scattering by a heavy target, nearly at rest. We will find potential corrections, linear in this ratio, associated with both the next-to-eikonal expansion and, at the same level, corrections due to the nonlinear gravitational interactions.

In section \ref{seceik} we will review the calculation for the leading eikonal phase to establish our conventions. We study $n$-graviton exchange, and the self-consistency of expansion around the eikonal limit in Sec.\ \ref{neik}, going on to review the exponentiation of the leading eikonal.   We proceed to calculating the next-to-eikonal power correction associated with expanding the light scalar propagator in one-loop diagrams in section \ref{noneik}.  At the same power, one-loop diagrams with seagull vertices and graviton trees are calculated in section \ref{sec:gravtree}.    In section \ref{sec:multi-grav} we embed these one loop corrections in ladder exchange at arbitrary order, and find that the result is consistent with the exponentiation of the first power correction.   In four dimensions, only the gravitational correction survives, and we calculate the correction to the saddlepoint at this order.   We conclude with a summary and brief discussion of our results. 


\section{Relativistic Scattering}
\label{secrel}

It was shown by `t Hooft \cite{tHooft:1987vrq} that the scattering of ultrarelativistic particles could be reliably studied by considering graviton exchange in perturbative quantum gravity. In the rest frame of one of the particles, the gravitational field of the rapidly moving particle is that of a gravitational shock wave, which can be described by the Aichelburg-Sexl metric \cite{Aichelburg:1970dh}. The particle at rest is a quantum particle whose dynamics can be described by the solutions of Klein-Gordon equation in the Aichelburg-Sexl background. This was `t Hooft's approach \cite{tHooft:1987vrq} and the results can also be obtained by summing a class of Feynman diagrams in the eikonal approximation \cite{Kabat:1992tb,Giddings:2011xs}. In this section we will demonstrate how this latter approach can be extended to the next-to-eikonal level.

We will investigate the small-angle gravitational scattering of an ultra-relativistic light scalar particle of energy $E_{\phi}$ off of a very heavy scalar particle. The mass of the heavy particle, $M_{\sigma}$, is bigger than $E_{\phi}$ and both are much larger than the transferred momentum. This approximation 
is made not only because it
simplifies the calculations needed to determine the next-to-eikonal corrections.  It will also lead to power corrections of the form ${\Delta \over E_\phi}$, which in terms of invariants is $\sqrt{(-t)}M_\sigma/ (s-M_\sigma^2)$,  much larger than the corrections that characterize the scattering of equal-mass particles.

We first review the exponentiation of the amplitude in the leading eikonal approximation in the kinematic regime described above. Then we consider the next-to-leading eikonal corrections. 

To make our kinematics explicit, consider the scattering
\begin{align}
p\ + \ q\ &\rightarrow\ p'\ +\ q' \, ,\ p^2\ =\ p'{}^2\ =0\, , \ q^2\ =\ q'{}^2\ =\ M_\sigma^2\, ,
\end{align}
with
\begin{align}
\Delta\ &\equiv\ \sqrt{-(p'-p)^2} \ \ll\ E_\phi \equiv p^0\ \ll\ M_\sigma\, .
\end{align}
Thus, we take the incoming and outgoing momentum $p$ and $p'$ of the light scalar particle $\phi$ to be much larger than the momentum transfer from the heavy to the light scalar, $\Delta=p'-p$. We will take $q$ and $q'$ to be the incoming and outgoing momenta of the heavy scalar, labelled $\sigma$.  We will always work to leading power in $M_\sigma$, and seek the first power corrections in $E_\phi$.  Our calculations will be carried out in the de Donder gauge.   

For our calculations we choose to work in the symmetric frame where $\Delta^0=\Delta^z=0$, $p^0=p'^0\equiv E_\phi=|{\bf p}|$, which we take nearly in the z-direction.  In this frame we have
\bea
p^\mu\ &=&\ \left ( E_\phi, p^z,  -\frac{{\bf \Delta}}{2} \right)\, , \nonumber \\
p'{}^\mu\ &=&\ \left ( E_\phi, p^z, + \frac{{\bf \Delta}}{2} \right)\, ,
\label{eq:kinematics}
\eea
with $p^z=p'{}^z=\sqrt{E_\phi^2-\Delta^2/4}$.  Here, $p^z=E_\phi$ up to corrections of relative order $E_\phi^{-2}$, which we may neglect through the first nonleading power. 

\subsection{Eikonal Phase}
\label{seceik}

Let us first consider this process in the eikonal limit to establish our conventions. In this limit it is well known that only ladder and crossed-ladder diagrams contribute. It will be instructive to treat the cases where one and two gravitons are  exchanged between the scalars before moving to the general $n$-graviton case.

\subsubsection{One Graviton Exchange}

\begin{figure}
\includegraphics[width=2.5in]{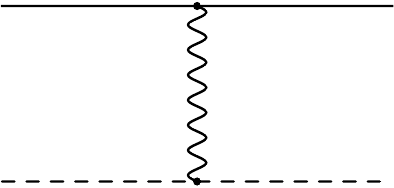}
\caption{A scattering process with a single graviton exchanged. The heavy scalar is the solid line and the light scalar is the dotted line.}
\label{fig1}
\end{figure}

Let us first work with the simplest case where only a single graviton is exchanged (Fig.\ \ref{fig1}). The matrix element corresponding to this diagram can be written as
\begin{align}
i\mathcal{M}^0_1&=(-\frac{i\kappa}{2})^2\frac{i}{2}\frac{L_{\mu\nu\alpha\beta}}{\Delta^2+i\epsilon}
\tau^{\alpha\beta}(p,p',M_\phi)\tau^{\mu\nu}(q,q',M_\sigma)\, ,
\end{align}
where
\begin{align}
L_{\mu\nu\alpha\beta}=\eta_{\mu\alpha}\eta_{\nu\beta}+\eta_{\mu\beta}\eta_{\nu\alpha}-\eta_{\mu\nu}\eta_{\alpha\beta}
\end{align}
is the numerator of the de Donder gauge graviton propagator, and
\begin{align}
\tau^{\mu\nu}(q,q',M_\sigma)=q^\mu q'^\nu+q^\nu q'^\mu-\eta^{\mu\nu}((q\cdot q')-M_\sigma^2)
\label{eq:q-numerator}
\end{align}
is the scalar-scalar-graviton vertex, and $\kappa^2=32\pi G$.   For amplitudes ${\cal M}_n^j$, here and below subscripts label the number of gravitons attached to the heavy scalar, and superscripts the power in $\Delta/E_\phi$, so that superscript 0 corresponds to the eikonal approximation.

We will use the fact that in the large $M_\sigma$ limit
\begin{align}
L_{\mu\nu\alpha\beta}\tau^{\mu\nu}(q,q',M_\sigma)=2M_\sigma^2(2\delta_{0\alpha}\delta_{0\beta}-\eta_{\alpha\beta})+\mathcal{O}(M_\sigma\Delta)\, .
\label{LargeM}
\end{align}
To leading power in $M_\sigma$, the right hand side of \eqref{LargeM} is independent of the momentum flowing through the vertex so we will always be free to use this identity for any vertex on a heavy line.

In the ultra-relativistic limit $M_\phi=0$ and $p^0={\bf |p|} = E_\phi$ so,
\begin{align}
(2\delta_{0\alpha}\delta_{0\beta}-\eta_{\alpha\beta})\tau^{\alpha\beta}(p,p',M_\phi)=4E_\phi^2+\mathcal{O}(\Delta^2)\, .
\label{zero}
\end{align}
Corrections are down by two powers of $E_\phi$.
In the $\Delta^0=0$ frame, $p^0=p'^0$  by construction, and \eqref{zero}  holds.   

We emphasize that the terms we have ignored at leading eikonal order in \eqref{LargeM} and \eqref{zero} are suppressed either by a factor of $M_\sigma$ and $E_\phi^2$ respectively. Thus, when we consider the $\frac{1}{E_\phi}$ next-to-eikonal contribution in the following discussion, we will not need to consider corrections arising from such numerator factors. This is one of the primary benefits of working in this particular kinematic regime.

Using \eqref{LargeM} and \eqref{zero}, we have
\begin{align}
i\mathcal{M}^0_1=(-\frac{i\kappa}{2})^2\frac{i}{2}(2M_\sigma^2)(4E_\phi^2)\frac{1}{\Delta^2+i\epsilon}\, .
\end{align}

\subsubsection{Two Graviton Exchange}

\begin{figure}
\includegraphics[width=2.5in]{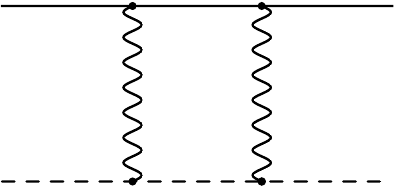}
\caption{A two-graviton ladder diagram. The heavy scalar is the solid line and the light scalar is the dotted line.}
\label{fig2}
\end{figure}

\begin{figure}
\includegraphics[width=2.5in]{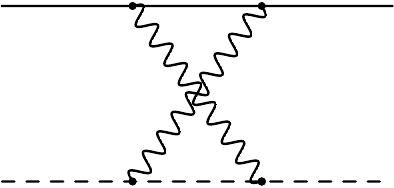}
\caption{A two-graviton crossed ladder diagram. The heavy scalar is the solid line and the light scalar is the dotted line.}
\label{fig3}
\end{figure}

Let us now consider the scattering process where two gravitons are exchanged. In this case the matrix element becomes
\begin{align}
i\mathcal{M}^0_2&=(-\frac{i\kappa}{2})^4(\frac{i}{2})^2(i)^2(2\pi)^{-4}(2M_\sigma^2)^2
\nonumber \\
&\times\int d^4k_1 d^4k_2\delta^4(k_1+k_2+\Delta)
\nonumber \\
&\times \tau^{\alpha_1\beta_1}(p,p-k_1,M_\phi)\tau^{\alpha_2\beta_2}(p-k_1,p',M_\phi)
\nonumber \\
&\times\frac{2\delta_{0\alpha_1}\delta_{0\beta_1}-\eta_{\alpha_1\beta_1}}{k_1^2+i\epsilon}\frac{2\delta_{0\alpha_2}\delta_{0\beta_2}-\eta_{\alpha_2\beta_2}}{k_2^2+i\epsilon}\frac{1}{(p-k_1)^2+i\epsilon}
\nonumber \\
&\times \left [\frac{1}{(q+k_1)^2-M_\sigma^2+i\epsilon}+\frac{1}{(q+k_2)^2-M_\sigma^2+i\epsilon} \right ]\, ,
\label{2grav1}
\end{align}
where we have made use of \eqref{LargeM}. Note that the sum in the last line of \eqref{2grav1} corresponds to combining the ladder (Fig.\ \ref{fig2}) and crossed ladder (Fig.\ \ref{fig3}) diagrams. In the large $M_\sigma$ limit and the $\Delta^0=0$ frame, we have,
\begin{align}
&\delta(k_1^0+k_2^0+\Delta^0)
\nonumber \\ &
\times \left [\frac{1}{(q+k_1)^2-M_\sigma^2+i\epsilon}+\frac{1}{(q+k_2)^2-M_\sigma^2+i\epsilon} \right ]
\nonumber \\
&=(2M_\sigma)^{-1}\delta(k_1^0+k_2^0) \left [\frac{1}{k_1^0+i\epsilon}+\frac{1}{k_2^0+i\epsilon} \right ]\, .
\label{preiden}
\end{align}
This expression can be simplified by use of the identities (see \cite{Saotome:2012vy} and references therein), which we will have several occasions to use below, on both the heavy and light scalar lines.
\begin{align}
&\sum_{\text{Perms of }\omega_i}\delta(\omega_1+...+\omega_n)
\frac{1}{\omega_{1}+i\epsilon}...\frac{1}{\omega_{1}+...+\omega_{n-1}+i\epsilon}
\nonumber \\
& = \delta(\omega_1+...+\omega_n)\ \sum_{\omega_n} \prod_{i=1}^{n-1} \frac{1}{\omega_j+i\epsilon}
\nonumber\\
&=(-2\pi i)^{n-1}\delta(\omega_1)...\delta(\omega_n)\, .
\label{iden}
\end{align}
In \eqref{2grav1}, $\omega_i=k_i^0$, $n=2$. After applying \eqref{iden} we have,
\begin{align}
i\mathcal{M}^0_2&=(-\frac{i\kappa}{2})^4(\frac{i}{2})^2(i)^2(2\pi)^{-4}M_\sigma(-2\pi i)
\nonumber \\
&\times\int d^4k_1 d^4k_2\delta(k_1^0)\delta(k_2^0)\delta^3({\bf k_1+k_2+\Delta})
\nonumber \\
&\times \tau^{\alpha_1\beta_1}(p,p-k_1,M_\phi)\tau^{\alpha_2\beta_2}(p-k_1,p',M_\phi)
\nonumber \\
&\hspace{-5mm}  \times\frac{2\delta_{0\alpha_1}\delta_{0\beta_1}-\eta_{\alpha_1\beta_1}}{k_1^2+i\epsilon}\frac{2\delta_{0\alpha_2}\delta_{0\beta_2}-\eta_{\alpha_2\beta_2}}{k_2^2+i\epsilon}\frac{1}{(p-k_1)^2+i\epsilon}\, .
\end{align}
Note that the delta functions over $k_1^0$ and $k_2^0$ guarantee that $p^0=(p^0-k_1^0)=p'^0$ so we are free to use the light scalar vertex identity, \eqref{zero}. After applying this identity ${\mathcal M}^0_2$ becomes
\begin{align}
i\mathcal{M}^0_2&=(-\frac{i\kappa}{2})^4(\frac{i}{2})^2(i)^2(2\pi)^{-4}M_\sigma(-2\pi i)(4E_\phi^2)^2
\nonumber \\
&\times\int d^4k_1 d^4k_2\delta(k_1^0)\delta(k_2^0)\delta^3({\bf k}_1+{\bf k}_2+{\bf \Delta})
\nonumber \\
& \times\frac{1}{k_1^2+i\epsilon}\frac{1}{k_2^2+i\epsilon}\frac{1}{(p-k_1)^2+i\epsilon}\, .
\end{align}
We next go on to extend results of this form to arbitrary numbers of gravitons.

\subsection{$n$-Graviton Exchange and Expansion Around the Eikonal Limit}
\label{neik}

Let us now consider the form of the matrix element when $n$ gravitons are exchanged. In this case, the matrix element is

\begin{align}
i\mathcal{M}^0_n&=(-\frac{i\kappa}{2})^{2n}(\frac{i}{2})^n(i)^{2n-2}(2\pi)^{4-4n}(2M_\sigma^2)^n
\nonumber \\
&\times\int d^4k_1...d^4k_n\delta^4(k_1+...+k_n+\Delta)
\nonumber \\
&\times\prod_{i=1}^n[\tau_i^{\alpha_i\beta_i}(p-K_{i-1},p-K_i,M_\phi)
 \frac{2\delta_{0\alpha_i}\delta_{0\beta_i}-\eta_{\alpha_i\beta_i}}{k_i^2+i\epsilon}]
 \nonumber \\
&\times\prod_{i=1}^{n-1}\frac{1}{(p-K_i)^2+i\epsilon}
\nonumber \\
&\times\sum_{\text{perms of }k_i} \left [\frac{1}{(q+k_1)^2-M_\sigma^2+i\epsilon} \right.
\nonumber \\
& \left.\times...\frac{1}{(q+k_1+...+k_{n-1})^2-M_\sigma^2+i\epsilon} \right ]\, ,
\label{ngrav1}
\end{align}
where $k_i$ are the graviton momenta and 
\begin{align}
K_i\equiv\sum_j^i k_j\, .
\label{eq:K_i-def}
\end{align}
 The tensors $\tau_i^{\alpha_i\beta_i}(p-K_{i-1},p-K_i,M_\phi)$ are the $n$ scalar-scalar-graviton vertices on the light scalar line. We have made use of the heavy scalar vertex approximation, Eq.\ \eqref{LargeM}. As we saw in the two graviton case, the summation over all permutations of the ordering of the graviton momenta onto the heavy line generates all the diagrams.

Just as in the two graviton case, we can simplify \eqref{ngrav1} by applying the identities \eqref{iden} and \eqref{zero} in succession. The result is
\begin{align}
i\mathcal{M}^0_n&=(-\frac{i\kappa}{2})^{2n}(\frac{i}{2})^n(i)^{2n-2}(2\pi)^{4-4n}
\nonumber \\
&\times 2M_\sigma^{n+1}(-2\pi i)^{n-1}(4E_\phi^2)^n
\nonumber \\
&\times\int d^4k_1...d^4k_n\delta(k_1^0)...\delta(k_n^0)\delta^3({{\bf k}_1+...+{\bf k}_n+\Delta})
\nonumber \\
&\times\prod_{i=1}^n\frac{1}{k_i^2+i\epsilon}
\prod_{i=1}^{n-1}\frac{1}{(p-K_i)^2+i\epsilon}
\nonumber \\
&=-2i(2\pi)^3M_\sigma \left (-\frac{\kappa^2M_\sigma E_\phi^2}{2(2\pi)^3} \right )^n
\nonumber \\
&\times\int d^3{\bf k_1}...d^3{\bf k_n}\, \delta^3({\bf k}_1+...+ {\bf k}_n+ {\bf \Delta})
\nonumber \\
&\times\prod_{i=1}^n\frac{1}{\bf k_i^2}
\prod_{i=1}^{n-1}\frac{1}{2{\bf p \cdot K}_i-{\bf K}_i^2+i\epsilon}\, .
\label{norig}
\end{align}
In this frame, the scalar propagators in Eq.\ (\ref{ngrav1}) can be expanded to the second inverse power of $E_\phi$ as
\begin{eqnarray}
\frac{1}{2{{\bf p} \cdot {\bf K}_i}-{\bf K}_i^2+i\epsilon}
&=& \frac{1}{2E_\phi K^z_i - {\bf \Delta}\cdot {\bf K}^\perp_i - {\bf K}_i^2 +i\epsilon}
\nonumber \\
&\ & \hspace{-33mm} \approx \frac{1}{2E_\phi} \left (\frac{1}{K_i^z+i\epsilon}+\frac{ ({\bf \Delta}+{\bf K}^\perp_i )\cdot {\bf K}^\perp_i+ K_i^z{}\, ^2}{2E_\phi\, (K_i^z+i\epsilon)^2} \right)\, ,
\nonumber\\
\label{eq:prop1}
\end{eqnarray}
where $K_i$ is defined in Eq.\ (\ref{eq:K_i-def}).

For the leading power eikonal phase we will only need to keep the first term in the expansion.  For the expansion to be meaningful, however, we must confirm that the ratios  $({\bf \Delta}+{\bf K}^\perp_i )\cdot {\bf K}^\perp_i/E_\phi K_i^z$ and $K^z_i{}{\, ^2}/E_\phi K_i^z$ are small.   This is trivially the case for the term with $K_i^z{}^2$, but for the transverse components we have to take into account that the integration contour for each $z$-component $k_i^z$ of loop momentum $i$ can be pinched between poles from the graviton propagators through which it flows.  This possibility can already be seen for the $n=2$ ladder and crossed ladder, Figs.\ \ref{fig2} and \ref{fig3}.  In both cases, in addition to the single pole from the $p-k$ line, there are two pairs of poles in the $k^z$ plane, at 
\begin{equation}
k^z\ =\ \pm i | \bf{k}^\perp |
\end{equation}
and at 
\begin{equation}
k^z\ =\ \pm i |  \bf{\Delta}+ \bf{k}^\perp |\, .
\end{equation}
These pairs of poles each pinch the $k^z$ integral, forcing the $k_z$ contour to pass through regions where its size is of order of magnitude of $|\bf{k}^\perp|$ or of 
$|\bf{\Delta}+\bf{k}^\perp|$.   These quantities may be much smaller than $|\bf{\Delta}|$, and indeed may vanish.   In the specific expansion of Eq.\ (\ref{eq:prop1}), however, the numerator is linear in both ${\bf k}^\perp$ and ${\bf \Delta}+\bf{k}^\perp$, ensuring that the ratio of this  correction term is always small.   This shows that the expansion is self-consistent for $n=2$.   

In fact, we can extend this argument, and show the self-consistency of the expansion  for all $n$.
The $j$th graviton propagator provides poles  located at $k_j^z=\pm i{\bf k}^\perp_j$, which pinch the $k^z_j$ contour at zero when transverse components vanish.  The expansion still makes sense, however, if we can find at least one momentum $k_j^z$, $j\le i$, which contributes to $K_i^z$ and which can be deformed away from the origin to order $\Delta$.   That is, we only need to find one loop momentum $k_j^z$, $j\le i$ that flows to the heavy mass propagator and back on lines that both carry transverse momenta of order $\Delta$.   Suppose no such loop exists.   This would require that every ${\bf k}_j^\perp=0$, for all $j\le i$ or for all $j>i$.    This would impliy in turn that either ${\bf K}_i^\perp=0$ or ${\bf \Delta}+{\bf K}^\perp_i=0$.    

In summary, whenever the numerators of the ratios in which we are expanding are nonzero, we can find an integration contour along which the denominators, $E_\phi K_i^z$ , remain larger than the numerators.  This includes those regions where $K_i^z$ is pinched at zero, because those pinches require the numerator to vanish even faster.    An expansion in ${\bf K}_i^\perp\cdot ({\bf \Delta}+{\bf K}_i^\perp)/2E_\phi K_i^z$ therefore makes sense.   Notice that this would not have been the case if we had expanded in, for example, ${\bf K}_i^\perp{}^2/E_\phi K_i^z$ alone.  

For now, we limit ourselves to the leading eikonal phase from the first term of the expansion in Eq.\ (\ref{eq:prop1}).   We will return to the treatment of nonleading terms, such as those in (\ref{eq:prop1}) in the following section.  

It will be useful to symmetrize across the $n$ graviton momenta so that we have
\begin{align}
i\mathcal{M}^0_n&=-4i(2\pi)^3E_{\phi}M_\sigma\frac{1}{n!} \left(-\frac{\kappa^2M_\sigma E_\phi}{4(2\pi)^3} \right)^n
\nonumber \\\, 
&\times\int d^3{\bf k_1}...d^3{\bf k_n}\, \delta^3({\bf k_1+...+k_n+\Delta})
\nonumber \\
&\times\prod_{i=1}^n\frac{1}{\bf k_i^2}
\sum_{\text{perms of }k_i^z} \left [\frac{1}{k_1^z+i\epsilon}...\frac{1}{k_1^z+...+k_{n-1}^z+i\epsilon} \right ]\, .
\label{eq:symm-n}
\end{align}
In this form, we can again apply \eqref{iden}, this time for the $k^z$ components, to get
\begin{align}
i\mathcal{M}^0_n&=4(2\pi)^2E_\phi M_\sigma\frac{1}{n!} \left (i\frac{\kappa^2M_\sigma E_\phi}{4(2\pi)^2} \right)^n\int d^3{\bf k_1}...d^3{\bf k_n}
\nonumber \\
&\times\delta(k_1^z)...\delta(k_n^z)\delta^2({\bf k_1^\perp+...+k_n^\perp+\Delta^\perp})
\prod_{i=1}^n\frac{1}{\bf k_i^2}\, .
\label{eq:M_n0_3k}
\end{align}
Now let us Fourier transform into (transverse) impact parameter space,
\begin{align}
i\widetilde{ \mathcal{M}}^0_n({\bf b}^\perp) &= \int\frac{d^2{\bf \Delta} }{(2\pi)^2}e^{i{\bf b}^\perp\cdot {\bf \Delta}^\perp}
\mathcal{M}^0_n({\bf \Delta}^\perp)
\nonumber\\
&=4(E_\phi M_\sigma)\frac{1}{n!} \left (i\frac{\kappa^2M_\sigma E_\phi}{4(2\pi)^2} \right)^n\int d^2{\bf k_1^\perp}...d^2{\bf k_n^\perp}
\nonumber \\
& \hspace{10mm}\times
\prod_{i=1}^n \left [\frac{1}{(\bf k_i^\perp)^2}e^{-i{\bf b^\perp\cdot k_i^\perp}} \right ]\, .
\label{eq:chi-0-n-b}
\end{align}
After summing over all $n$ we have,
\begin{align}
i\widetilde{\mathcal{M}}^0=2(s- M_\sigma^2)(e^{i\chi_0}-1)\, .
\label{eq:M0-exp}
\end{align}
The integrals in Eq.\ (\ref{eq:chi-0-n-b}) for $\chi_0$ are divergent in four dimensions.   Whenever necessary we shall evaluate such integrals $D=4-2\epsilon$ dimensions. 
With this regularization, $\chi_0$ is given by 
\begin{align}
\tilde \chi_0\left({\bf b}^\perp\right)&=\frac{\kappa^2M_\sigma E_\phi}{4(2\pi)^{2-2\epsilon}}\int d^{2-2\epsilon}{\bf k^\perp}\frac{1}{(\bf k^\perp)^2}e^{-i{\bf b^\perp\cdot k^\perp}}
\nonumber \\[2mm]
&=2GM_\sigma E_\phi \ b^{2\epsilon}\, \pi^{\epsilon}\, \Gamma(-\epsilon) \, .
\label{eq:chi0}
\end{align}
As is clear from the form of the integral, the regularization is necessary in the infrared, with $\epsilon<0$.
Here and below, the required integrals and properties of special functions can all be found in Ref.\ \cite{GradRy}.
Expanding around $\epsilon=0$, we have
\begin{align}
\chi_0\left({\bf b}^\perp\right)&=4GM_\sigma E_\phi \bigg [\frac{1}{d-4}-\log b
\nonumber\\
&  \hspace{10mm} +\ \text{terms independent of }b \bigg ]\, ,
\label{chi0}
\end{align}
 where $b=|{\bf b}|$.
Below we will encounter a related integral with an additional (positive) term in the denominator,
\bea
&\ & \frac{1}{(2\pi)^{2-2\epsilon}}\int d^{2-2\epsilon}{{\bf k}^\perp}\frac{1}{({\bf k}^\perp)^2+(k^z)^2}e^{-i{\bf b\cdot k^\perp}}
\nonumber\\
&\ & \hspace{5mm} =\ b^{2\epsilon}  \left[ 
 \frac{(k^zb)^{-\epsilon}}{(2\pi)^{1-\epsilon}}\ K_{-\epsilon}\left( k^z b \right)\, \right]\, .
 \label{eq:K-integral}
 \eea
 The function $\chi_0$ in Eq.\ (\ref{eq:chi0}) can also be found from the $k^z\to 0$ limit of this expression.

The inverse Fourier transform of (\ref{chi0}) back to momentum space is dominated by the stationary phase point at
\begin{align}
b^\perp\sim\frac{4GM_\sigma E_\phi}{\Delta} = 2 R_s {E_{\phi} \over \Delta^{\perp}}\, ,
\label{saddle_b1}
\end{align}
where $R_s$ is the Schwarzschild radius of the heavy particle and ${\Delta}^2=t$.
Thus in the ultrarelativistic regime of small momentum transfer, this process is dominated by  impact parameters 
larger than the Schwarzschild radius of the target particle.  We note, however, that because of the asymmetry between the masses of the two scalar particles, the point of stationary phase of the inverse transform occurs at a multiple of the Schwarzschild radius that is set by the ratio of the momentum transfer to the energy, not to the heavy particle mass.   Therefore, at small but finite angles, the scattering is dominated by impact parameters of order $R_s$ divided by the scattering angle, rather than by the ratio of the momentum transfer divided by the total center of mass energy.  In this situation, non-linear gravitational effects and corrections to the eikonal approximation are of comparable size.   We will return the transformation to impact parameter below, when interpreting next-to-eikonal corrections, neglecting any contributions that are concentrated at $b^\perp=0$.

\subsection{Next-to-Eikonal Power at One Loop}
\label{noneik}

\begin{figure}
\includegraphics[width=2.5in]{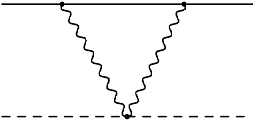}
\caption{A diagram with a seagull interaction. The heavy scalar is the solid line and the light scalar is the dotted line.}
\label{fig4}
\end{figure}

\begin{figure}
\includegraphics[width=2.5in]{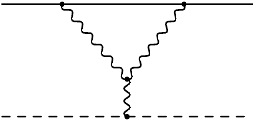}
\caption{A diagram with a triangle interaction. The heavy scalar is the solid line and the light scalar is the dotted line.}
\label{fig5}
\end{figure}

There are many types of corrections that have been studied to the set of diagrams discussed in the previous section \cite{Amati:1987wq, Giddings:2011xs}. We will be only considering the following three types of related corrections of the same order in $\Delta/E_\phi$. First, we include those coming from the next term to the scalar propagator in the eikonal approximation. In other words, for each scalar in turn, we will keep at most one insertion of the second term in the propagator expansion of equation \eqref{eq:prop1}.   The other two corrections are shown in (Fig.\ \ref{fig4}) and (Fig.\ \ref{fig5}) which are also inserted only once.   One-loop corrections with three graviton-scalar vertices on the light or heavy scalar line, and self energies of the graviton give no corrections of the form $\Delta/E_\phi$.   (As we have seen, there are no $\Delta/E_\phi$ corrections arising from the numerator factors of single-graviton exchange, which also simplifies our calculations.)  We call this the next-to-eikonal approximation, including the specifically gravitational corrections.  We will show that next to eikonal corrections are consistent with exponentiation in impact parameter space, and we determine the leading corrections to the eikonal phase ($\chi_0$) of the previous section. 

Let us first consider the contribution that arises when we retain the next order term in the expansion of the light scalar propagators, Eq.\ \eqref{eq:prop1}. The correction to the amplitude in Eq.\ \eqref{norig} is then, 
\begin{align}
i\mathcal{M}_2^{1a}({\bf \Delta})&=-2i(2\pi)^{3-2\epsilon}M_\sigma \left (-\frac{\kappa^2M_\sigma E_\phi^2}{2(2\pi)^{3-2\epsilon}} \right)^2(2E_\phi)^{-2}
\nonumber \\
& \hspace{-8mm}\times\int d^3{\bf k_1}d^3{\bf k_2}\delta^3({\bf k_1+k_2+\Delta})
\frac{1}{\bf k_1^2}\frac{1}{\bf k_2^2}
\frac{  {\bf \Delta}\cdot {\bf k}_1^\perp + {\bf k}_1\, {}^2}{(k_1^z+i\epsilon)^2}\, .
\nonumber\\
 & \hspace{-10mm} =\ 2i(2\pi)^{3-2\epsilon}M_\sigma \left (-\frac{\kappa^2M_\sigma E_\phi^2}{2(2\pi)^{3-2\epsilon}} \right )^2\, \frac{1}{4E_\phi^2}
\nonumber \\
& \hspace{-5mm} \times \int \frac{d^{3-2\epsilon}{\bf k}_1}{{\bf k}_1^2} \frac{d^{3-2\epsilon}{\bf k}_2}{{\bf k}_2^2}\; \delta^{2-2\epsilon}\left( {\bf \Delta}+\sum_{l=1}^2 {\bf k}^\perp_l\right)
\nonumber\\
& \hspace{-5mm} \times\  \left[ \left( {\bf k}^\perp_1\cdot {\bf k}^\perp_2 - \left( k^z_1 \right)^2 \right)\frac{1}{\left( k^z_1 \right)^2} \right ] \delta\left( k^z_1+k^z_2 \right)\, ,
\label{eq:M-nlp-n=2}
\end{align}
where as above, we define the integrals by dimensional regularization with $\epsilon<0$.
The Fourier transform to impact parameter space is 
\bea
{\cal M}^{1a}_2 ({\bf b})\ &=&\ 2i\, \left( s - M_\sigma^2\right) \ \frac{\kappa^4 M^2_\sigma E_\phi}{32(2\pi)^{5-4\epsilon}} 
\nonumber \\
&\ & \hspace{-10mm} \times \int \frac{dk_1^z}{\left( k^z_1 \right)^2}\
 \int d^{2-2\epsilon}{\bf k}_1 \frac{ e^{-i{\bf k}^\perp_1\cdot {\bf b}}}{({\bf k}_1^\perp)^2+ (k^z_1)^2}
 \nonumber\\
 &\ & \hspace{-10mm} \times\
 \int d^{2-2\epsilon}{\bf k}_2\, \frac{ e^{-i{\bf k}^\perp_2\cdot {\bf b}}}{({\bf k}_2^\perp)^2+ (k^z_1)^2}
\nonumber\\
&\ & \hspace{-5mm} \times\  \left( {\bf k}^\perp_1\cdot {\bf k}^\perp_2 - \left( k^z_1 \right)^2 \right)
\nonumber\\
&\equiv& 2i(s-M_\sigma^2)\ \chi^{1a}_2  \, ,
\label{eq:nlp-explicit-bspace}
\eea
where, to isolate a potential contribution to the phase, we introduce the notation $\chi^{1a}_2$, which will appear again at higher order.
The transverse integrals are given by
\bea
&\ & \frac{1}{(2\pi)^{2-2\epsilon}}\int d^{2-2\epsilon}{\bf k^\perp}\frac{1}{({\bf k}^\perp)^2+(k^z)^2}\,  k_i^\perp \,e^{-i{\bf b^\perp\cdot k^\perp}}
\nonumber\\
&\ & \hspace{5mm} =\ \frac{ib^{2\epsilon-2}}{(2\pi)^{1-\epsilon}}\, b_i^\perp 
\big [ 2\epsilon\, 
 (k^zb^\perp)^{-\epsilon}\, K_{-\epsilon}\left( k^z b^\perp \right)
 \nonumber\\
 &\ & \hspace{15mm}
 -\ (k^zb^\perp)^{1-\epsilon}\, K_{1+\epsilon}\left( k^z b^\perp \right)\, 
 \big ]\, .
 \label{eq:K-integral-deriv}
 \eea
This result can be derived from (\ref{eq:K-integral}), using the functional relation
 \bea
  \frac{d}{dz}\ \left(z^\nu K_\nu(z)\right)\ =\ -\, z^{\nu}K_{\nu-1}(z)\, ,
 \eea
 and the invariance of these Bessel functions under a sign change in their arguments: $K_\mu(z)=K_{-\mu}(z)$.
The resulting integrals over products of Bessel functions are then given by (note in this form the explicit symmetry under changes in sign of the Bessel function indices)
\bea
\int_0^\infty \frac{dz}{z^\lambda}\ K_\mu(z)\, K_\nu(z)\ &=&\ 
\nonumber\\
&\ & \hspace{-30mm} \frac{2^{-\lambda-2}}{\Gamma(1-\lambda)} \
\prod_{\eta,\zeta=\pm 1} \ \Gamma\left( \frac{1 - \lambda + \eta \mu + \zeta \nu}{2}\right)\, .
\label{eq:GR-for-Ks}
\eea
In spite of the $(k^z)^{-2}$ singularity, the $z$ integrals encountered in the evaluation of Eq.\ (\ref{eq:nlp-explicit-bspace}) are finite at the origin in the $k^z$ plane in $D>4$ dimesnions once the transverse integrals are carried out.  This may be seen from an expansion of the resulting Bessel functions in Eq.\ (\ref{eq:K-integral-deriv}).   

Applying these integrals to Eq.\ (\ref{eq:nlp-explicit-bspace}), the amplitude is found to be
\bea
{\cal M}^{1a}_2 ({\bf b})\ &=&\  2i\, \left( s - M_\sigma^2\right) \ \frac{\kappa^4 M^2_\sigma E_\phi}{32(2\pi)^{3-2\epsilon}} 
\nonumber \\
&\ & \hspace{-10mm} \times b^{4\epsilon-1}\, 2^{-2\epsilon}\ \frac{ \Gamma\left(\frac{1}{2} \right) \; \Gamma^2\left(\frac{1}{2} -\epsilon\right)\;
\Gamma \left(\frac{1}{2} -2\epsilon\right)}{\Gamma(-2\epsilon)}
\nonumber\\
&\equiv& 2i\left(s-M_\sigma^2\right)\, \tilde\chi_2^a({\bf b})
\, ,
\nonumber\\
\label{eq:nlp-chi1a}
\eea
which gives an explicit form for the function $\chi_2^a(b)$ as a function of $\epsilon$.
This result  is consistent in analytic structure in the $\epsilon$ plane with the one-loop next-to-leading power found in the analysis of string-brane scattering in  Ref.\ \cite{DAppollonio:2010krb}\footnote{The variable $p$ there is to be interpreted as $6+2\epsilon$ in this comparison.}.   In particular, it vanishes in four dimensions.   It potentially provides, however, a finite correction starting at two loops, given precisely by this result times the divergent single-graviton phase in impact parameter space.   

\subsection{Graviton trees at one loop}
\label{sec:gravtree}

Let us now consider the graviton tree corrections, starting with the seagull interaction, Fig.\ \ref{fig4}, which we denote as
\begin{align}
i\mathcal{M}_2^{1sg}&=(-\frac{i\kappa}{2})^2(i\kappa^2)(\frac{i}{2})^2(i)(2\pi)^{-4}(2M_\sigma^2)^2(\frac{1}{2})
\nonumber \\
&\times\int d^4k_1 d^4k_2\delta^4(k_1+k_2+\Delta)\tau^{\alpha_1\beta_1\alpha_2\beta_2}(p,p')
\nonumber \\
&\times\frac{2\delta_{0\alpha_1}\delta_{0\beta_1}-\eta_{\alpha_1\beta_1}}{k_1^2+i\epsilon}\frac{2\delta_{0\alpha_2}\delta_{0\beta_2}-\eta_{\alpha_2\beta_2}}{k_2^2+i\epsilon}
\nonumber \\
&\times \left [\frac{1}{(q+k_1)^2-M_\sigma^2+i\epsilon}+\frac{1}{(q+k_2)^2-M_\sigma^2+i\epsilon} \right ]\, .
\label{eq:seagull0}
\end{align}
We have again symmetrized graviton momenta and used the heavy scalar vertex approximation, Eq.\ \eqref{LargeM}. 
The explicit scalar-graviton seagull vertex, $\tau^{\alpha_1\beta_1\alpha_2\beta_2}(p,p')$ in this expression, can be found in the convenient appendices of Refs.\ \cite{BjerrumBohr:2002ks} and \cite{BjerrumBohr:2004mz}.

We now use the identities \eqref{preiden} and \eqref{iden} to simplify Eq.\ \eqref{eq:seagull0}, 
\begin{align}
i\mathcal{M}_2^{1sg}&=(-\frac{i\kappa}{2})^2(i\kappa^2)(\frac{i}{2})^2(i)(2\pi)^{-4}(2M_\sigma^2)^2(\frac{1}{2})
\nonumber \\
&\times(2M_\sigma)^{-1}(-2\pi i)\int d^4k_1 d^4k_2\delta(k_1^0)\delta(k_2^0)
\nonumber \\
&\times\delta^3({\bf k}_1+{\bf k}_2+{\bf \Delta})\tau^{\alpha_1\beta_1\alpha_2\beta_2}(p,p')
\nonumber \\
&\times\frac{2\delta_{0\alpha_1}\delta_{0\beta_1}-\eta_{\alpha_1\beta_1}}{k_1^2+i\epsilon}\frac{2\delta_{0\alpha_2}\delta_{0\beta_2}-\eta_{\alpha_2\beta_2}}{k_2^2+i\epsilon}\, .
\label{eq:seagull}
\end{align}
The numerator factors for the seagull diagram are readily evaluated, and give
\begin{align}
&(2\delta_{0\alpha_1}\delta_{0\beta_1}-\eta_{\alpha_1\beta_1})(2\delta_{0\alpha_2}\delta_{0\beta_2}-\eta_{\alpha_2\beta_2})\tau^{\alpha_1\beta_1\alpha_2\beta_2}(p,p')
\nonumber \\
&=(4p^0p'^0-p\cdot p')\approx4E_\phi^2\, ,
\label{eq:seagull-num}
\end{align}
where we have used that fact that $p\cdot p'$ is subleading by two powers of $E_\phi$ compared to $p^0p'^0$ in the ultra-relativistic limit. 

Organizing its factors in the same manner as for the seagull, the graviton triangle diagram, Fig.\ \ref{fig5} is given by 
\begin{align}
i\mathcal{M}_2^{1tri}&=(-\frac{i\kappa}{2})^4 (\frac{i}{2})^3(i)(2\pi)^{-4}(2M_\sigma^2)^2(\frac{1}{2})
\nonumber \\
&\times(2M_\sigma)^{-1}(-2\pi i)\int d^4k_1 d^4k_2\delta(k_1^0)\delta(k_2^0)
\nonumber \\
&\times\delta^3({\bf k}_1+{\bf k}_2+{\bf \Delta}) \tau^{\mu\nu}(p,p')\ \frac{L_{\mu\nu\eta\lambda}}{{\bf \Delta}^2}\ 
\nonumber \\
&\times\frac{2\delta_{0\alpha_1}\delta_{0\beta_1}-\eta_{\alpha_1\beta_1}}{k_1^2+i\epsilon}\
\frac{2\delta_{0\alpha_2}\delta_{0\beta_2}-\eta_{\alpha_2\beta_2}}{k_2^2+i\epsilon}
\nonumber \\
& \times \tau^{\alpha_1\beta_1\alpha_2\beta_2\eta\lambda}(k_1,k_2)
 \, ,
\label{eq:triangle}
\end{align}
where $\tau^{\alpha_1\beta_1\alpha_2\beta_2\eta\lambda}(k_1,k_2)$ is the three graviton vertex, the moderately lengthy expression for which may be found in \cite{BjerrumBohr:2002ks,BjerrumBohr:2004mz}.

The relevant numerator factors for the triangle diagram are found by a straightforward calculation, and gives
\begin{align}
&(2\delta_{0\alpha_1}\delta_{0\beta_1}-\eta_{\alpha_1\beta_1})(2\delta_{0\alpha_2}\delta_{0\beta_2}-\eta_{\alpha_2\beta_2})
\nonumber \\
&\times \tau^{\mu\nu}(p,p')\, L_{\mu\nu\eta\lambda}\tau^{\alpha_1\beta_1\alpha_2\beta_2\eta\lambda}(k_1,k_2)
\nonumber \\
&= 8 E_\phi^2 \left( k_1^z{}^2 + k_2^z{}^2 \right)\ + \cdots\, .
\label{eq:triangle-num}
\end{align}
 Terms not explicitly included are either suppressed by the ratio $p\cdot p'/E_\phi^2$ or are proportional to $k_i^2$, $i=1,2$.   The former are negligible to first non-leading power, while the second give integrals that vanish in dimensional regularization.   
 
 Applying Eqs.\ \eqref{eq:seagull} -- \eqref{eq:triangle-num},  we find for the sum of the seagull and triangle diagrams,
\begin{align}
i\mathcal{M}_2^{1b}({\bf \Delta})&= 8i M_\sigma (2\pi)^3 \left( \frac{-\kappa^2M_\sigma E_\phi}{4(2\pi)^3} \right)^2 (\frac{1}{2})
\nonumber \\
& \hspace{-10mm} \times \int d^3{\bf k_1}d^3{\bf k_2}
\delta^3({{\bf k}_1+{\bf k}_2+\Delta})\frac{1}{{\bf k}_1^2}\frac{1}{{\bf k}_2^2}\; \frac{\frac{1}{2}{\bf \Delta}^2 + k_1^z{}^2}{{\bf \Delta}^2}
\nonumber\\
&= 2(S-M_\sigma^2) i \chi^{b}_2\left(\Delta^\perp\right)
\, .
\label{seapart}
\end{align}
where in the second equality we isolate the tree-level prefactor.  This correction to the amplitude in \eqref{seapart} has the same power behavior in $E_\phi$ and  $\Delta$ as the lowest-order power corrections given above in Eq.\ \eqref{eq:M-nlp-n=2}.   The dimensionless factor that multiplies the prefactor is
\begin{align}
i\chi_2^b\left(\Delta^\perp\right) &=   i\; \left( \frac{\kappa^4 M_\sigma^2 E_\phi}{16(2\pi)^3} \right)\ 
\nonumber\\
& \hspace{-10mm}\times \int d^3{\bf k_1}d^3{\bf k_2}
\delta^3({\bf k}_1+{\bf k}_2+{\bf \Delta})\frac{1}{{\bf k}_1^2}\frac{1}{{\bf k}_2^2}\; \frac{{\bf \Delta}^2 + \frac{1}{2}k_1^z{}^2}{{\bf \Delta}^2}\, .
\end{align}
These integrals are not difficult to evaluate, \footnote{ We note that they occur as leading terms in scalar mass in the calculations of Appendix B in Ref.\ \cite{BjerrumBohr:2002ks}.}    using dimensional regularization in terms of which the relevant integrals are finite \cite{BjerrumBohr:2002ks,Donoghue:1994dn}. The result in four dimensions is,
\begin{align}
i\chi_2^b\left(\Delta^\perp\right) &=  i\; \left( \frac{\kappa^4 M_\sigma^2 E_\phi}{16(2\pi)^3} \right)\ \frac{15\pi^3}{16}\ \frac{1}{\Delta^\perp}\, .
\label{eq:chi2b-eval}
\end{align}
   For use below, we  transform  $\chi_2^{b}$ to impact parameter space,
\begin{align}
i\widetilde{ \chi}_2^{b}({\bf b}^\perp) &= i\; \int\frac{d^2{\bf \Delta} }{(2\pi)^2}e^{i{\bf b}^\perp\cdot {\bf \Delta}^\perp}
\nonumber\\
 &= i\;  \left( \frac{\kappa^4 M_\sigma^2 E_\phi}{256\pi} \right)\ \frac{15}{16}\ \frac{1}{b^\perp}\, .
\label{eq:chi21b-trans}
\end{align}
We will encounter this contribution to the imaginary part in our discussion of $n$-graviton exchange in the next subsection.  

Finally, we note that the diagrams that are mirror reflections of those in (Fig.\ \ref{fig4}) and (Fig.\ \ref{fig5}) (i.e., those with the two distinct matter-graviton vertices on the light line) are suppressed in the limit of a heavy sigma particle. This simplifies the combinatorics of the next section and is a practical advantage of this particular kinematic limit.


\section{The first power correction to multi-graviton exchange}
\label{sec:multi-grav}

In the following, we extend our discussion to the first power correction encountered in diagrams with ladder exchange at arbitrary order.   We will see that these corrections are consistent with the same exponentiation of the leading phase found from pure ladders in the eikonal approximation.

\subsection{$n$-Graviton Ladders}

\begin{figure}
\includegraphics[width=3in]{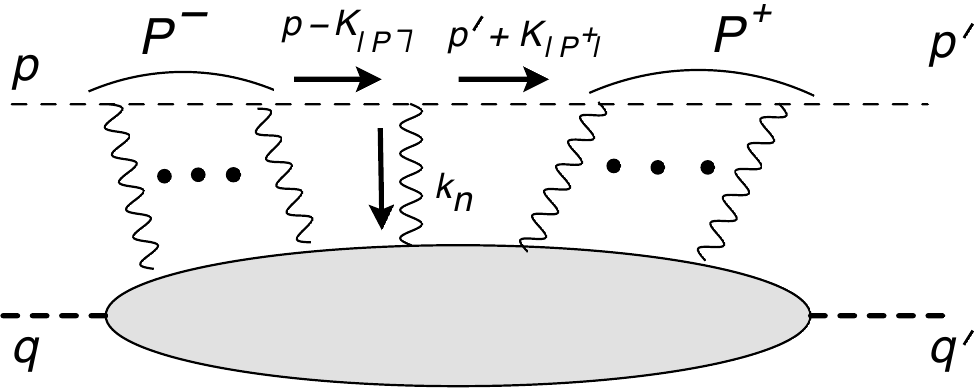}
\caption{Flow of momentum for a general permutation of graviton lines connected to the light scalar in Eq.\ (\ref{eq:M-all-perms}).}
\label{fig:Pplusminus}
\end{figure}

Let us consider the set of all $n$-graviton ladder diagrams, organized as in Eq.\ (\ref{eq:nlp-explicit-bspace}), but not yet in the eikonal approximation. 
We reorganize all possible permutations of the $\{k_i\}$ into sets $P^-$ and $P^+$ of lines attached to the light scalar line before and after the graviton carrying momentum $k_n$, respecively, with explicit permutations of all lines within $P^\pm$.   
This flow of momentum is illustrated in Fig.\ \ref{fig:Pplusminus}.

We denote by $|P^\pm|$ the number of elements in each set.  Graviton momenta in $|{\cal P}^-|$ flow  into the light scalar $(p)$ line from the line labelled $k_n$ in the direction of the initial state, and $|{\cal P}^+|$ momenta flow from $k_n$ onto the light scalar in  the direction of the final state. 
\begin{align}
i\mathcal{M}_n&=2(2\pi)^3M_\sigma \left (i\frac{\kappa^2M_\sigma E_\phi^2}{2(2\pi)^3} \right )^n\ \frac{1}{n!}
\nonumber \\
&\times\int d^3{\bf k}_1...d^3 {\bf k}_n\, \delta^3({\bf k}_1+...+{\bf k}_n+\Delta)\, \prod_{i=1}^n\frac{1}{{\bf k}_i^2}
\nonumber \\
&\times
\sum_{ \{P^\pm\} }\, \Bigg \{ \sum_{\mathrm{perms}\, k_b\in P^+}\, \left[ \prod_{b=1}^{|P^+|}\frac{i}{ (p'+K_b)^2+i\epsilon} \right]
\nonumber\\
& \times \ \sum_{\mathrm{perms}\, k_a\in P^-}\, \left[ \prod_{a=1}^{|P^-|}\frac{i}{ (p-K_a)^2+i\epsilon} \right] \Bigg \}\, ,
\label{eq:M-all-perms}
\end{align}
with the partial sums $K_{a,b}$ defined as in Eq.\ (\ref{eq:K_i-def}) for each permutation of gravitons within the sets $P^\pm$.  Note the momenta of lines in set $P^+$ appear in light scalar propagators with the momentum of the outgoing light scalar, $p'$.

In the frame of Eq.\ (\ref{eq:kinematics}) for momenta $p^\mu$ and $p'\, {}^\mu$, we can rewrite Eq.\ (\ref{eq:M-all-perms}) as 
\begin{align}
i\mathcal{M}_n&=4(2\pi)^3M_\sigma E_\phi \left (i\frac{\kappa^2M_\sigma E_\phi}{4(2\pi)^3} \right )^n\ \frac{1}{n!}
\nonumber \\
&\times\int d^3{\bf k}_1...d^3 {\bf k}_n\, \delta^3({\bf k}_1+...+{\bf k}_n+\Delta)\, \prod_{i=1}^n\frac{1}{{\bf k}_i^2}
\nonumber \\
& \hspace{-5mm}\times
\sum_{ \{P^\pm\} }\, \Bigg \{ \sum_{\mathrm{perms}\, k_b\in P^+}\, \left[ \prod_{b=1}^{|P^+|}
\frac{i}{ - K_b^z - \left( \frac{ {\bf \Delta}\cdot {\bf K}^\perp_b+\vec{\bf K}_b^2}{2E_\phi}   \right )+i\epsilon} \right ]
\nonumber\\
& \hspace{-5mm} \times \ \sum_{\mathrm{perms}\, k_a\in P^-}\, \left[ \prod_{a=1}^{|P^-|}
\frac{i}{  K_a^z - \left( \frac{ {\bf \Delta}\cdot {\bf K}^\perp_a+\vec{\bf K}_a^2}{2E_\phi}   \right )+i\epsilon} \right] \Bigg \}\, .
\label{eq:M-all-perms-Ks}
\end{align}
The expansion of each denominator of this expression as in Eq.\ (\ref{eq:prop1}) gives the full next-to-leading power result for these diagrams.   In the notation we have adopted, we write
\begin{align}
i\mathcal{M}_n\ =\ i\mathcal{M}^0_n\ +\ i\mathcal{M}^{1a}_n\ +\ \dots
\label{eq:M-to-nlp}
\end{align}
where   $\mathcal{M}^0_n$ is given in Eq.\ (\ref{eq:M_n0_3k}) and where $\mathcal{M}^{1a}_n$ summarizes all linear expansions, which are manifestly suppressed by one power of $E_\phi$.   Further corrections are suppressed by higher powers of $\Delta/E_\phi$.    
We now show how the leading corrections may be organized to provide a compact result, which generalizes the one-loop calculations of the previous section.

  In organizing the first power correction, we make use of the feature that any term explicitly proportional to $\vec {\bf k}_i^2$, for any graviton momentum $\vec {\bf k}_i$, results in a delta function $\delta^{2-2\epsilon}({\bf b})$ in impact parameter space.   We neglect all such contributions, since we are interested here in finite-$b$ behavior.   With this in mind, The term we are after, $ i\mathcal{M}^{1a}_n$ is the sum of contributions from terms that are linear in invariants ${\bf \Delta}\cdot {\bf k}_i$, and inner products $\vec{\bf k}_\alpha \cdot \vec{\bf k}_\beta$  from the squared partial sums ${\bf K}_i^2$ in Eq.\ (\ref{eq:M-to-nlp}).    Our notation for these corrections will be
\bea
i{\cal M}^{1a}_n({\bf \Delta}) \ &=&\ 2(2\pi)^{3-2\epsilon} M_\sigma \left( \frac{i\kappa^2M_\sigma E_\phi}{4(2\pi)^{3-2\epsilon}}\right)^n \, \frac{1}{n!}
\nonumber\\
&\ & \times \int \prod_{i=1}^n\ \frac{d^{3-2\epsilon}{\bf k}_i}{{\bf k}_i^2}\
\delta^{3-2\epsilon}({\bf \Delta}+\sum_{i=1}^n {\bf k}_i )\, 
\nonumber\\
&\ & \times \left[ \sum_{i=1}^n\ {\cal J}_i^{(n)}\ + \ \sum_{\{\alpha,\beta\}} {\cal J}_{\alpha\beta}^{(n)} \right]\, .
\label{eq:cal-J-sum}
\eea
We have introduced dimensional regularization to ensure that our integrals are well-defined.
We will deal with the two sets of ${\cal J}$s in (\ref{eq:cal-J-sum}) in turn, starting with $\{{\cal J}_i\}$.  Each ${\cal J}_i$ collects all terms proportional to ${\bf \Delta}\cdot \vec{\bf k}_i={\bf \Delta}\cdot {\bf k}^\perp_i$, where we recall that in the frame we are using, $\Delta^z=0$.   These terms can be written as
\bea
 {\cal J}_i^{(n)}\ &=&\  {\bf \Delta}^\perp\cdot {\bf k}^\perp_i\ \sum_{\{P^\pm\}}
 \nonumber\\
 &\ & \times \Bigg \{ 
\sum_{\text{Perm}\{k_b\} \in P^+} \prod_{b=1}^{|P^+|}\frac{i}{ - K_b^z+i\epsilon}\sum_i^{|P^+|}\frac{ \theta_{i,b}}{ - K_b^z+i\epsilon} 
\nonumber\\
&\ & \hspace{5mm} \times\ \sum_{\text{Perm}\{k_a\} \in P^-} \prod_{a=1}^{|P^-|}\frac{i}{ K_a^z+i\epsilon}
\nonumber\\
&\ & +\ 
\sum_{\text{Perm}\{k_b\} \in P^+} \prod_{b=1}^{|P^+|}\frac{i}{ - K_b^z+i\epsilon} 
\nonumber\\
&\ & \hspace{5mm} \times\ \sum_{\text{Perm}\{k_a\} \in P^-} \prod_{a=1}^{|P^-|}\frac{i}{ K_a^z+i\epsilon}
\sum_i^{|P^-|}\frac{ \theta_{i,a}}{  K_a^z+i\epsilon} 
\Bigg \} \, ,
\nonumber\\
\label{eq:T-i-sum}
\eea
where $\theta_{ia}$ is one if momentum $k_i$ is carried by line $a$, and is zero otherwise.  This result can be rewritten as a derivative with respect to $k_i^z$, as
 \bea
 {\cal J}_i^{(n)}\ &=&\   {\bf \Delta}^\perp\cdot {\bf k}^\perp_i\  
 \nonumber\\
 &\ & \hspace{-5mm} \times\
 \epsilon_i\frac{\partial}{\partial k^z_i}\  \sum_{ \{P^+\}\{P^-\}  }\ \Bigg \{ \sum_{\text{Perm}\{k_b\} \in P^+} \prod_{b=1}^{|P^+|}\frac{i}{ - K_b^z+i\epsilon} 
\nonumber\\
&\ & \hspace{-5mm} \times\ \sum_{\text{Perm}\{k_a\} \in P^-} \prod_{a=1}^{|P^-|}\frac{i}{ K_a^z+i\epsilon} \Bigg \}\, ,
 \label{eq:T_i-def}
 \eea
 where the sign function $\epsilon_i$ is defined to provide a relative minus sign for $k_i$, depending on whether it is included in a partition of lines that attach to the light scalar before or after the graviton with momentum $k_n$,
 \bea
 \epsilon_i\ &=&\ +1\, , \quad k_i\ \in\ P^+\, ,
\nonumber\\
 \epsilon_i\ &=&\ -1\, , \quad k_i\ \in\ P^-\, .
 \eea
 We now apply a fundamental identity for eikonal sums, which is closely related to the result, Eq.\ (\ref{iden}), used above for the purely eikonal approximation, 
 \begin{align}
 \sum_{\text{Perm}\{k_b\} \in P^+} \prod_{b=1}^{|P^+|}\frac{i}{ - K_b^z+i\epsilon}
 &=
 \prod_{b=1}^{|P^+|}\  \frac{i}{-k_b^z + i\epsilon }
 \nonumber\\
 \sum_{\text{Perm}\{k_a\} \in P^-} \prod_{b=1}^{|P^+|}\frac{i}{  K_a^z+i\epsilon}
 &=
 \prod_{a=1}^{|P^+|}\  \frac{i}{ k_a^z + i\epsilon }\, .
\label{eq:eikonal-fact}
 \end{align}
 We review a proof of this eikonal identity because we will use the technique in our treatment of ${\cal J}_{\alpha\beta}^{(n)}$.  We rewrite the product of eikonal denominators associated with a specific permutation of the lines in $P^+$ as
\begin{align}
 \prod_{b=1}^{|P^+|}\frac{i}{ - K_b^z+i\epsilon}
 &=
 \prod_{b=1}^{|P^+|}\  \int_{x_{b-1}}^\infty dx_b\, e^{-i \left (k_b^z - i\epsilon \right )x_b}\, ,
 \label{eq:eik-to-x-ordered}
 \end{align}
with $x_0\equiv 0$.  A similar form is easily found for the $P^-$ denominators.   If we now sum over all permutations, the $x_b$ integrals decouple, and we find
\begin{align}
\sum_{\text{Perm}\{k_b\} \in P^+} \frac{i}{ - K_b^z+i\epsilon}
&= \prod_{b=1}^{|P^+|}\  \int_{0}^\infty dx_b\, e^{-i \left (k_b^z - i\epsilon \right )x_b}
\nonumber\\
& = \prod_{b=1}^{|P^+|}\ \frac{i}{-k_b^z+i\epsilon}\, ,
\label{eq:eikonal-to-x-all}
\end{align}
in which dependence is fully factorized. 
 
 The identities (\ref{eq:eikonal-fact})  completely factor the $k_i^z$ dependences for ${\cal J}_i^{(n)}$ in Eq.\ (\ref{eq:T_i-def}), and for arbitrary $n$ we find,
 \bea
 {\cal J}_i^{(n)}\ &=& \ {\bf \Delta}\cdot {\bf k}_i^\perp\
 \nonumber\\
 &\ & \ \times \epsilon_i\frac{\partial}{\partial k^z_i}\,  \sum_{ \{P^+\}\{P^-\}  }\  \prod_{b=1}^{|P^+|}\  \frac{i}{-k_b^z + i\epsilon }
\ \prod_{a=1}^{|P^-|}  \frac{i}{k_a^z + i\epsilon }
\nonumber\\
&=&\ {\bf \Delta}^\perp\cdot{\bf k}^\perp_i\ \left[ \frac{i}{(-k_i^z+i\epsilon)^2} + \frac{i}{(k_i^z+i\epsilon)^2} \right]
\nonumber\\
&\ & \times\ \sum_{ \{P^+\}\{P^-\}/k_i  }\  \prod_{b=1}^{|P^+|}\  \frac{i}{-k_b^z + i\epsilon }
\ \prod_{a=1}^{|P^-|}  \frac{i}{k_a^z + i\epsilon }
\nonumber\\
&=&\ 2{\bf \Delta}^\perp\cdot{\bf k}^\perp_i\ \left[ \frac{i}{(k_i^z)^2} \right]\ 
(2\pi)^{n-2}\ \prod_{j=1,\ne i}^{n-1}\; \delta \left(k_j^z\right)\, .
\nonumber\\
\label{eq:T_i-full-integrals}
\eea
where in the second expression, $\{P^+\}\{P^-\}/k_i$ denotes  partitions without the graviton of momentum $k_i$.  As we shall see below, the double pole at $k^z_i=0$ will be regulated dimensionally, and no $i\epsilon$ prescription will be necessary.
This form is clearly well-organized for combination with a transform to impact parameter space, and shows that in general the evaluation of the resulting expressions will require dimensional continuation.   All the remaining $k_j^z$ dependence reduces to delta functions, as in the leading eikonal amplitude, because the remaining sum over partitions provides the combination $i/(-k_j^z+i\epsilon)+i/(k_j^z+i\epsilon)$ for each of the remaining $k_j\ne k_i$.

   For the ${\cal J}_{\alpha\beta}^{(n)}$, terms, linear in the three-dimensional scalar products ${\bf k}_\alpha\cdot {\bf k}_\beta$, we can begin with the same steps, starting with the analog of Eq.\ (\ref{eq:T-i-sum}).
   \bea
 {\cal J}^{(n)}_{\alpha\beta}\ &=&\  \vec{\bf k}_\alpha\cdot \vec{\bf k}_\beta\ \sum_{P^+,P^-}
 \nonumber\\
 &\ & \times \Bigg \{ 
\sum_{\text{Perm}\{k_b\} \in P^+} \prod_{b=1}^{|P^+|}\frac{i}{ - K_b^z+i\epsilon}\sum_b^{|P^+|}\frac{ \theta_{(\alpha\beta),b}}{ - K_b^z+i\epsilon} 
\nonumber\\
&\ & \hspace{5mm} \times\ \sum_{\text{Perm}\{k_a\} \in P^-} \prod_{a=1}^{|P^-|}\frac{i}{ K_a^z+i\epsilon}
\nonumber\\
&\ & +\ 
\sum_{\text{Perm}\{k_b\} \in P^+} \prod_{b=1}^{|P^+|}\frac{i}{ - K_b^z+i\epsilon} 
\nonumber\\
&\ & \hspace{0mm} \times\ \sum_{\text{Perm}\{k_a\} \in P^-} \prod_{a=1}^{|P^-|}\frac{i}{ K_a^z+i\epsilon}
\sum_a^{|P^-|}\frac{ \theta_{(\alpha\beta),a}}{ - K_a^z+i\epsilon} 
\Bigg \}\, ,
\nonumber\\
\label{eq:T-alpha-beta-sum}
\eea
where here the step functions $\theta_{(\alpha\beta),c}$ restrict the sum to lines that carry both of the momenta $k_\alpha$ and $k_\beta$.   We will now give a form in which this expansion can be rewritten in terms of a momentum derivative, as we did for ${\cal J}_i^{(n)}$.   

The appropriate derivative depends on the ordering of the gravitons of momenta $k_\alpha$ and $k_\beta$ along the light scalar line.
Consider first those ladder permutations in which $\alpha>\beta$, that is, momentum $k_\beta$ flows out of the light scalar line before $k_\alpha$.   The ${\bf k}_\alpha\cdot {\bf k}_\beta$ terms are generated by derivatives with respect to $k_\beta$, because every line that carries $k_\beta$ also carries $k_\alpha$.   The case when $\beta>\alpha$ is clearly reversed, and all diagrams fall into one of the two cases.   We shall refer to these sets of permutations within $P^+$ as $P^+_{\alpha>\beta}$ and $P^+_{\beta>\alpha}$.

We can separate the two sums over diagrams with $\alpha>\beta$ and $\beta>\alpha$ by inserting a step function in the integral representation of the product of eikonal denominators, Eq.\ (\ref{eq:eik-to-x-ordered}).   
\begin{align}
P^+_{\alpha>\beta}& \equiv \prod_{b=1,\alpha>\beta}^{|P^+|}\; \frac{i}{ - K_b^z+i\epsilon}
\nonumber\\
 &= \prod_{b=1}^{|P^+|}\  \int_{x_{b-1}}^\infty dx_b\, e^{-i \left (k_b^z - i\epsilon \right )x_b}\, \theta(x_\alpha-x_\beta) 
 \label{eq:eik-to-x-alpha-beta}
 \end{align}
 where the second relation imposes the constraint on the eikonal products.
Summing over all permutations subject  only to this step function, we factor $k_\alpha$ and $k_\beta$ from the remaining momenta in $P^+$, which
themselves completely factorize as above.   For convenience of notation, we represent the result as
\begin{align}
 { \cal P}^+_{\alpha>\beta} &\equiv \sum_{\mathrm perms} P^+_{\alpha>\beta} 
\nonumber\\
& \hspace{-5mm} =\ \prod_{b=1}^{|P^+|}\  \int_0^\infty dx_b\, e^{-i \left (k_b^z - i\epsilon \right )x_b}\ \theta(x_\alpha-x_\beta)
\nonumber\\
& \hspace{-5mm}  = \frac{i}{(- k_\alpha -k_\beta+i\epsilon )}\ \frac{i}{(-k_\alpha+i\epsilon)}\ \left( \prod_{b=1\ne \alpha,\beta}^{|P^+|}\ \frac{i}{-k_b+i\epsilon} \right)
\, .
\label{eq:cal-P-plus-alpha-beta}
\end{align}
To this result, we will apply a derivative with respect to $k_\beta^z$ to derive the corresponding terms in ${\cal J}_{\alpha\beta}^{(n)}$, Eq.\ (\ref{eq:T-alpha-beta-sum}).
The remaining terms contribute a result, ${\cal P}^+_{\beta>\alpha}$, which simply reverses the roles of $\alpha$ and $\beta$, and to which a derivative with respect to $k_\alpha^z$ is applied.

The same reasoning applies, of course, to the products of eikonal denominators for gravitons in each set $P^-$, deriving in the same manner  the quantity
\begin{align}
 { \cal P}^-_{\alpha>\beta} &\equiv \sum_{\mathrm perms} P^-_{\alpha>\beta} 
\nonumber\\
& \hspace{-5mm} =\ \prod_{a=1}^{|P^-|}\  \int_{-\infty}^0 dx_b\, e^{-i \left (k_b^z + i\epsilon \right )x_a}\ \theta(x_\alpha-x_\beta)
\nonumber\\
& \hspace{-5mm}  = \frac{i}{(k_\alpha +k_\beta+i\epsilon )}\frac{i}{(k_\beta+i\epsilon)}\ \left( \prod_{a=1\ne \alpha,\beta}^{|P^-|}\ \frac{i}{k_a+i\epsilon} \right)
\, .
\label{eq:cal-P-minus-alpha-beta}
\end{align}
For this term, the derivative with respect to $k_\alpha^z$ will be applied.   Putting all the terms together, we find
  \bea
  {\cal J}^{(n)}_{\alpha\beta}\ &=&\  \vec{\bf k}_\alpha\cdot \vec{\bf k}_\beta 
  \Bigg \{ \left [  \frac{\partial}{\partial k_\beta}\, {\cal P}^+_{\alpha>\beta}\ +\ \frac{\partial}{\partial k_\alpha  }\, {\cal P}^+_{\beta>\alpha} \right ]\, {\cal P}^-
  \nonumber\\
&\ & \hspace{10mm} - \ {\cal P}^+\, \left [  \frac{\partial}{\partial k_\beta}\, {\cal P}^-_{\alpha>\beta}\ +\ \frac{\partial}{\partial k_\alpha  }\, {\cal P}^-_{\beta>\alpha} \right ]\, \Bigg \}\, ,
\nonumber\\
 \label{eq:cal-J-alpha-beta-derivative}
  \eea
  Where ${\cal P}^\pm$ represents the product of factorized denominators in the corresponding sets.
  Using the explicit expressions, (\ref{eq:cal-P-plus-alpha-beta}) and (\ref{eq:cal-P-minus-alpha-beta}), their analogs for $\beta>\alpha$,  and the eikonal identities (\ref{eq:eikonal-fact}),  Eq.\ (\ref{eq:cal-J-alpha-beta-derivative}) for ${\cal J}_{\alpha\beta}^{(n)}$ can be put a form where the derivatives can be carried out,
  \bea
  {\cal J}^{(n)}_{\alpha\beta} &=& \vec{\bf k}_\alpha\cdot \vec{\bf k}_\beta\ \Bigg \{ 
  \frac{\partial}{\partial k^z_\alpha}\ \left[ \frac{i}{- k^z_\alpha - k^z_\beta ,+ i\epsilon}
  \frac{i}{ - k^z_\beta + i\epsilon} \right]
  \nonumber\\
  &\ & \hspace{5mm}
+ \frac{\partial}{\partial k^z_\beta}\ \left[ \frac{i}{(- k^z_\alpha - k^z_\beta + i\epsilon)} \frac{i}{ (- k^z_\alpha + i\epsilon ) } \right ]
\nonumber\\
&\ & \hspace{5mm} - \frac{\partial}{\partial k^z_\alpha}\ \left[ \frac{i}{ (k^z_\alpha + k^z_\beta + i\epsilon) } \frac{i}{ ( k^z_\beta + i\epsilon ) } \right]
  \nonumber\\
  &\ & \hspace{5mm}
- \frac{\partial}{\partial k^z_\beta}\ \left[ \frac{i}{ (k^z_\alpha + k^z_\beta + i\epsilon) } \frac{i}{ ( k^z_\alpha + i\epsilon ) } \right ] \Bigg \}
\nonumber\\
&\ &  \times\ \sum_{ \{P^+\}\{P^-\}/\{k_\alpha k_\beta\} }\  \prod_{b=1}^{|P^+|}\  \frac{i}{-k_b^z + i\epsilon }
\ \prod_{a=1}^{|P^-|}  \frac{i}{k_a^z + i\epsilon }
\nonumber\\
&=&\ \vec{\bf k}_\alpha\cdot \vec{\bf k}_\beta \ (-2i)\, {\rm Im}\, \left[ \frac{1}{(k_\alpha+k_\beta+i\epsilon)}\frac{1}{(k_\alpha+i\epsilon)}\frac{1}{(k_\beta+i\epsilon)} \right]\ 
\nonumber\\
&\ & \times
(2\pi)^{n-3}\ \prod_{a=1,\ne \alpha,\beta }^{n-1}\; \delta \left(k_a^z\right)\, .
\label{eq:J_alpha-beta-full-integrals}
\eea
The factor with $k_\alpha$ and $k_\beta$ can be simplified further by the distribution identity,
\bea
&& \hspace{-4mm} -\ 2\, {\rm Im}\, \left[ \frac{1}{(k_\alpha+k_\beta+i\epsilon)}\frac{1}{(k_\alpha+i\epsilon)}\frac{1}{(k_\beta+i\epsilon)} \right]
\nonumber\\
&\ & \hspace{-2mm} =  2\pi\,  \left[ \delta\left ( k^z_\alpha \right) \frac{1}{(k^z_\beta)^2} + \delta\left ( k^z_\beta\right) \frac{1}{(k^z_\alpha)^2}
\ -\ \delta\left ( k^z_\alpha +k^z_\beta \right) \frac{1}{(k^z_\alpha)^2}\, \right ]\, .
\nonumber\\
\label{eq:-distribution-identity}
\eea
 When integrated against an even function of the $k^z_i$s, any double poles at the origin in this expression vanish, so there is no need to keep track of $i\epsilon$ terms in these denominators, which are defined by dimensional regularization.   
 
We are now ready to derive our general expression for the factorized next-to-leading power correction associated with the expansion of denominators.   We do this by using Eq.\ (\ref{eq:-distribution-identity}) in (\ref{eq:J_alpha-beta-full-integrals}) for ${\cal J}^{(n)}_{\alpha\beta}$ and combine the result with Eq.\ (\ref{eq:T_i-full-integrals}) for ${\cal J}^{(n)}_i$ in Eq.\ (\ref{eq:cal-J-sum}).  We then eliminate ${\bf \Delta}$ in favor of the ${\bf k}^\perp_l$, including $l=n$ by using momentum conservation.  This gives
\bea
i{\cal M}^{1a}_n({\bf \Delta}) \ &=&\ 2(2\pi)^{3-2\epsilon} M_\sigma \left( \frac{i\kappa^2M_\sigma E_\phi}{4(2\pi)^{3-2\epsilon}}\right)^n 
\nonumber\\
&\ & \hspace{-15mm} \times \int \prod_{i=1}^n\ \frac{d^{3-2\epsilon}{\bf k}_i}{{\bf k}_i^2}\
\delta^{3-2\epsilon}({\bf \Delta}+\sum_{i=1}^n {\bf k}_i )\, (2\pi)^{n-2}
\nonumber\\
&\ & \hspace{-15mm} \times\ \frac{i}{n!}\,  {\cal F}^{(n)}\left \{ k_i^z,{\bf k}_i^\perp \} \right)
\label{eq:M1a-calF}
\eea
where, expanding the three-dimensional scalar products in the ${\cal J}^{(n)}_{\alpha\beta}$ terms into transverse and $z$ components, we find a 
slightly complex result, labelled ${\cal F}^{(n)}$, which we organize further into a convenient form
\bea
{\cal F}^{(n)}\left \{ k_i^z,{\bf k}_i^\perp \} \right)\ &=&
\nonumber\\
&\ & \hspace{-30mm} -2 \sum_{i=1}^{n-1} \prod_{j=1,\ne i}^{n-1} \delta \left (k^z_j\right)
\left( {\bf k}^\perp_i{}^2 + \sum_{l =1,\ne i}^n {\bf k}^\perp_i\cdot {\bf k}^\perp_l \right) \, \frac{1}{(k^z_i)^2}
\nonumber \\
&\ & \hspace{-30mm} +\ \sum_{\{\alpha,\beta\}} \prod_{j=1,\ne \alpha,\beta}^{n-1} \delta \left (k^z_j\right) 
\left( {\bf k}^\perp_\alpha\cdot {\bf k}^\perp_\beta + k^z_\alpha k^z_\beta \right) 
\nonumber\\
&\ & \hspace{-30mm}
\times    \left ( \delta\left ( k^z_\beta\right) \frac{1}{(k^z_\alpha)^2} + \delta\left ( k^z_\alpha\right) \frac{1}{(k^z_\beta)^2}
\ -\ \delta\left ( k^z_\alpha+k^z_\beta \right) \frac{1}{(k^z_\alpha)^2}\, \right ) 
\nonumber\\
&\ & \hspace{-30mm} =\, -2 \sum_{i=1}^{n-1} \frac{{\bf k}_i^\perp\cdot {\bf k}_n^\perp }{(k_i^z)^2}\, \prod_{j\ne i}^{n-1} \delta(k_j^z) 
\nonumber\\
&\ & \hspace{-25mm}
-  \sum_{\{\alpha,\beta\}} \frac{ {\bf k}^\perp_\alpha\cdot {\bf k}^\perp_\beta}{(k_\alpha^z)^2}\, \delta(k_\alpha^z+k_\beta^z) \prod_{j=1\ne\alpha,\beta}^{n-1} \delta(k_j^z)
\nonumber\\
&\ &  \hspace{-25mm}  -2 \sum_{i=1}^{n-1}\ \sum_{l=1}^{n-1}\ \frac{{\bf k}_i^\perp\cdot {\bf k}_l^\perp }{(k_i^z)^2}\, \prod_{j\ne i}^{n-1} \delta(k_j^z) 
\nonumber\\
&\ & \hspace{-25mm} 
+  \sum_{\{\alpha,\beta\}} {\bf k}^\perp_\alpha\cdot {\bf k}^\perp_\beta  \left [ \frac{\delta(k_\beta^z)}{(k_\alpha^z)^2}\, +\,  \frac{\delta(k_\alpha^z)}{(k_\beta^z)^2} \right ] \prod_{j=1\ne\alpha,\beta}^{n-1} \delta(k_j^z)
\nonumber\\
&\ & \hspace{-25mm} +\ 2 \sum_{i=1}^{n-1}\, \prod_{j\ne i}^{n-1} \delta(k_j^z) 
\nonumber\\
&\ & \hspace{-25mm}
-  \sum_{\{\alpha,\beta\}} \frac{ k^z_\alpha  k^z_\beta}{(k_\alpha^z)^2}\, \delta(k_\alpha^z+k_\beta^z) \prod_{j=1\ne\alpha,\beta}^{n-1} \delta(k_j^z)\, .
 \label{eq:cal-F-def}
    \eea
    In the second equality, we have dropped terms that are identically zero, replaced ${\bf k}_i^\perp{}^2$ by $-(k_i^z)^2$, because a numerator factor of $\vec{\bf k}_i^2$ for any $i$ does not contribute at finite $b$, as noted above, and reorganized the sums.   The third and fourth sums, involving products of transverse vectors, cancel.
 Of the scalar products, those surviving are the first sum, invoving ${\bf k}_n\cdot {\bf k}_i$ when multiplied by $1/(k^z_i)^2$, and the second sum, over unordered pairs $\alpha,\beta$ that multiplies $\delta\left(k^z_\alpha +k^z_\beta\right)$.     There are $n-1$ terms of the former kind, for $i=1\dots n-1$, multiplied by 2, and $(n-1)(n-2)$ terms in the latter sum.   All terms in both sums are manifestly equal, up to changes of variable labels, giving an overall factor of $n(n-1)=n!/(n-2)!$ from this set of terms.  
  Similarly, the products $k^z_\alpha k^z_\beta$ in the sum over pairs either vanish or are equal to $-(k^z_\alpha)^2$.   All such terms with squared $z$ components again give the same result, and their counting is the same as for the transverse scalar products.   
  
  Using the cancellations and counting described above, the full result for the amplitude, Eq.\ (\ref{eq:M1a-calF}), in terms of an arbitrary choice of $k_\alpha$ and $k_\beta$ is now relatively simple, 
    \bea
i{\cal M}^{1a}_n({\bf \Delta}) \ &=&\ -2i(2\pi)^{3-2\epsilon} M_\sigma \left( \frac{i\kappa^2M_\sigma E_\phi}{4(2\pi)^{3-2\epsilon}}\right)^n 
\nonumber\\
&\ & \hspace{-15mm} \times \int \prod_{i=1}^n\ \frac{d^{3-2\epsilon}{\bf k}_i}{{\bf k}_i^2}\
\delta^{2-2\epsilon}({\bf \Delta}+\sum_{i=1}^n {\bf k}_i )
\nonumber\\
&\ & \hspace{-15mm} \times \frac{(2\pi)^{n-2}}{(n-2)!}\ \delta \left( \sum_{j=1}^n k^z_j\right)\ \prod_{j=1,\ne \alpha\beta}^{n-1} \delta \left (k^z_j\right)\ 
\nonumber\\
&\ & \hspace{-10mm} \times\  \left[ \left( {\bf k}^\perp_\alpha\cdot {\bf k}^\perp_\beta - \left( k^z_\alpha \right)^2 \right)\frac{1}{\left( k^z_\alpha \right)^2} \right ] \delta\left( k^z_\alpha+k^z_\beta \right)\, .
\nonumber\\
\label{eq:nlp-explicit}
\eea
For the special case, $n=2$, this agrees with ${\cal M}^{1a}_2$, the one-loop integral of Eq.\ (\ref{eq:M-nlp-n=2}).
Taking the transform to impact parameter space gives
 \bea
i\widetilde{ \mathcal{M}}_{n}^{1a}({\bf b}^\perp) \ &=& \ \int \frac{ d^{2-2\epsilon}{\bf \Delta}}{(2\pi)^{2-2\epsilon}}\ e^{i{\bf \Delta}\cdot {\bf b}}\ i{\cal M}^{1a}_n({\bf \Delta})
\nonumber\\
&\ & \hspace{-15mm} =\ -\frac{2}{(n-2)!}\frac{M_\sigma}{(2\pi)^{1-2\epsilon}} \left (i\frac{\kappa^2M_\sigma E_\phi}{4(2\pi)^{2-2\epsilon}}\right)^n
\nonumber \\
&\ & \hspace{-10mm} \times\ \left [\int \frac{d^2{\bf k^\perp}}{(k^\perp)^2}e^{-i{\bf b^\perp \cdot k^\perp}} \right ]^{n-2}
\nonumber \\
&\ & \hspace{-10mm} \times\ \left[   \int \frac{d^{3-2\epsilon}{\bf k_\alpha}}{{\bf k}_\alpha^2}\frac{d^{3-2\epsilon}{\bf k_\beta}}{{\bf k}_\beta^2}e^{-i({\bf k}^\perp_\alpha+{\bf k}^\perp_\beta)\cdot {\bf b}^\perp} \right.
\nonumber \\
&\ &\left. \hspace{-10mm} \times\ \left[ \left( {\bf k}^\perp_\alpha\cdot {\bf k}^\perp_\beta - \left( k^z_\alpha \right)^2 \right)\frac{1}{\left( k^z_\alpha \right)^2} \right ] \delta\left( k^z_\alpha+k^z_\beta \right)\right]\, .
\nonumber\\
\label{eq:new-basic-M1-0}
\eea
In terms of the lowest-order phase  $\tilde \chi_0$, given in Eq.\ (\ref{eq:chi0}), and the one-loop correction, $\chi^{1a}_2$ in Eq.\ (\ref{eq:nlp-chi1a}), this result is 
\bea
i\widetilde{ \mathcal{M}}_{n}^{1a}({\bf b}^\perp) 
&=& 
\nonumber\\
&\ & \hspace{-15mm} =\ \frac{2}{(n-2)!}(s-M_\sigma^2)(i\tilde \chi_0\left({\bf b}^\perp\right))^{n-2}(i\tilde \chi_2^{1a}\left({\bf b}^\perp\right))\, ,
\nonumber\\
\label{eq:new-basic-M1}
\eea
which shows that the first power correction found by expanding denominators appears times the expanded lowest-order phase at each order in the sum over ladders.
Although we suppress dependence on the dimensional parameter $\epsilon$, we recall that in four dimensions $\chi_2^{1a}$ vanishes linearly in $\epsilon$.    In general, of course, there are finite contributions to the amplitude from products of $\chi_2^{1a}$ with poles from $\chi_0$.   The result, Eq.\ (\ref{eq:new-basic-M1}) is consistent with exponentiation of this term at higher orders.  If $\chi_2^{1a}$ exponentiates, then real and imaginary parts generated by expanding the exponential to any fixed order in the absolute square of the amplitude will vanish.  These higher power considerations, however, are beyond this study.
 
\subsection{Combining a graviton tree with ladders}

We now go beyond purely ladder diagrams, and consider the sum of $n$-graviton exchange diagrams with a single seagull, which can be written as
\begin{align}
i\mathcal{M}_n^{1b\, sg}({\bf \Delta})&=\frac{(2\pi)^4}{E_\phi^2} \left (\frac{i\kappa^2M_\sigma^2 E_\phi^2}{(2\pi)^4} \right)^n\frac{1}{2}
\nonumber \\
&\times\int d^4k_1..d^4k_n\delta^4(k_1+..+k_n+\Delta)
\nonumber \\
&\times  \prod_{i=1}^{n}\frac{1}{k_i^2}\ \sum_{j=1}^{n-1}\ \prod_{\stackrel{i=1}{i\ne j}}^{n-1}\frac{1}{(p-K_i)^2+i\epsilon}
\nonumber \\
&\times \sum_{\text{perms of }k_i} \left [\frac{1}{(q+k_1)^2-M_\sigma^2+i\epsilon} \right.
\nonumber \\
& \left. \times...\frac{1}{(q+k_1+...+k_{n-1})^2-M_\sigma^2+i\epsilon} \right ]\, .
\label{eq:Mn1b}
\end{align}
The sum over index $j$ generates all places where the seagull vertex can be placed on the light scalar line, and the permutations over the $k_i$ generate all the diagrams for each placement of this vertex.   Note there is an over counting, because exchanging the order of the seagull legs does not result in distinct diagrams, so we divide by $2$. 
A similar manipulation can be carried out for the three-gluon triangle diagram.
The integrals of Eq.\ (\ref{eq:Mn1b}) and the corresponding expression where a three-gluon triangle replaces the seagull diagram, include the
same sum of permutations over $n-1$ heavy particle ($q$) propagators, with the same overall delta function
ensuring zero energy transfer.   
Thus we can
apply \eqref{iden} in the ultrarelativistic limit as we did for the purely eikonal exchange contribution.
On the light scalar line, there are now $n-1$ vertices and $n-2$ propagators.  Of these, $n-2$ vertices connect
to single gravitons, and at a single vertex to the sum of the seagull (Fig.\ \ref{fig4}) and graviton triangle diagrams (Fig.\ \ref{fig5}).   
In the notation of Eq.\ (\ref{eq:Mn1b}), the latter vertex is in position $j$, $1\le j \le n-1$, and carries momentum $k'_j\equiv k_j+k_{j+1}$ out
of the light scalar line.   Here, momenta $k_j$ and $k_{j+1}$ play the role that momenta $k_\alpha$ and $k_\beta$ played in our
analysis of $n$-single graviton exchange above.

The application of Eq.\ \eqref{iden} now factorizes the dependence of all of all single-graviton momenta on the heavy 
scalar line, setting all graviton energies to zero, and we find
\begin{align}
\label{eq:Mn1}
i\mathcal{M}_n^{1b}&=  \left (\frac{-\kappa^2M_\sigma E_\phi}{4(2\pi)^3} \right)^{n-2} 
\nonumber\\
& \times\ \sum_{j=1}^{n-1}\ \int d^3{\bf k}'_j\ i{\cal M}^{1b}_2\left({\bf k}'_j \right)
\nonumber \\
&\times\  \prod_{i=1}^{n}{}^{(j)}\int \frac{d^3{\bf k}_i}{{\bf k}^2_i} \delta^3( \sum_{i=1}^n{}^{(j)} {\bf k}_i+{\bf \Delta})
\nonumber \\
&\times   \ \prod_{i=1}^{n-1}{}^{(j)} \frac{1}{ K_i^z+i\epsilon}   
\, ,
\end{align}
where we absorb a three-dimensional delta function that sets ${\bf k}'_j={\bf k}_j+{\bf k}_{j+1}$ into the function,
${\mathcal M}_n^{1b}({\bf k}'_j)$, which then becomes the same function of momentum transfer  as in Eq.\ (\ref{seapart}) above, including its prefactor.
The superscript $(j)$ on the product reflects this change, and indicates that in the sums and products $k_j$ and $k_{j+1}$ are combined to a single term, $k'_j$.

We have in Eq.\ (\ref{eq:Mn1}) all placements of vertex $j$, which carries momentum ${\bf k}'_j$.  We may sum over 
all permutations of the $n-2$ remaining vertices on the $p$ line, which are all indistinguishable, and must be compensated 
for by an overall factor of $1/(n-2)!$.   We thus have
\begin{align}
i\mathcal{M}_n^{1b}&= \frac{1}{(n-2)!}\   \left (\frac{-\kappa^2M_\sigma E_\phi}{4(2\pi)^3} \right)^{n-2} 
\nonumber\\
& \times\ \int d^3{\bf k}'_j\ i{\cal M}^{1b}_2\left({\bf k}'_j\right)
\nonumber \\
&\times\  \prod_{\stackrel{i=1}{ i\ne j}}^{n-1}\int \frac{d^3{\bf k}_i}{{\bf k}^2_i} \delta^2( \sum_{i=1}^{n}{}^{(j)} {\bf k}_i+{\bf \Delta})
\nonumber \\
&  \times\   \delta( \sum_{i=1}^{n}{}^{(j)}k^z_i)\ \sum_{\stackrel{\text{perms of }}{k_i}}{}^{(j)} \prod_{j=1}^{n-1}{}^{(j)} \frac{1}{ K_j^z+i\epsilon}   \, .
\label{sean2e}
\end{align}
  For the set of momenta $k_1 \dots k_{j-1}, k'_j, k_{j+2} \dots k_n$, the final sum and product in this expression are in precisely the form necessary to apply the identity, Eq.\ \eqref{iden}, so that we can set all of their $z$ components to zero,
\begin{align}
i\mathcal{M}_n^{1b} \left({\bf \Delta}\right) &=\ \frac{1}{(n-2)!}\  \left (\frac{i\kappa^2M_\sigma E_\phi}{4(2\pi)^2} \right )^{n-2}  
\nonumber\\
& \times\ \int d^2{\bf k}'_j{}\ i{\cal M}^{1b}_2\left({\bf k}'_j{}\right)
\nonumber \\
&\times\  \prod_{\stackrel{i=1}{ i\ne j}}^{n-1}\int \frac{d^2{\bf k}_i}{{\bf k}^2_i} \delta^2( \sum_{i=1}^{n}{}^{(j)} {\bf k}_i+{\bf \Delta})\, .
\label{sean3e}
\end{align}
This expression is ready to be transformed to impact parameter space, which gives
\begin{align}
i\widetilde{\mathcal{M}}_n^{1b}\left({\bf b}\right) &=2(s- M_{\sigma}^2) \frac{(i\tilde \chi_0)^{n-2}}{(n-2)!}i \tilde \chi_2^{1b} \left({\bf b}\right)\, ,
\label{eq:M1nb}
\end{align}   
where the lowest-order example, $\widetilde {\mathcal M}^{1b}_2$ is the transform of the same combination of seagull and vertex diagrams given in Eq.\ (\ref{seapart}), and
where $\chi_2^{1b}$ is given in momentum and impact parameter space by Eqs.\ \eqref{eq:chi2b-eval} and \eqref{eq:chi21b-trans}, respectively.   

\subsection{The full next-to-leading power}
 
Summing over all $n$ in $i\widetilde{\mathcal{M}}^0+i\widetilde{\mathcal{M}}^1$, we can now combine the leading order result of
 Eq.\ \eqref{eq:M0-exp} with the first non leading powers from expanding light scalar denominators  \eqref{eq:new-basic-M1} which  begins at $n=2$,
  and with gravitational corrections \eqref{eq:M1nb}, which also begin at $n=2$, to find,
\begin{align}
i\widetilde{\mathcal{M}}^{0+1}\left({\bf b}\right) &=2(s- M_\sigma^2)
\nonumber\\
& \hspace{-10mm} \times\
\left[ 1+i\tilde{\chi}_2^{1b}\left({\bf b}\right)+i\tilde{\chi}_2^{1a}\left({\bf b}\right )\right ]e^{i\chi_0\left({\bf b}\right)} \, ,
\label{expon}
\end{align}
where again the power corrections are $i\tilde{\chi}_2^{1a}$ from expanding denominators and $i\tilde{\chi}_2^{1b}$from the seagull and graviton triangle.  Their explicit expressions are given in Eq.\ \eqref{eq:chi21b-trans} for $\tilde{\chi}_2^{1b}$ (in four dimensions) and \eqref{eq:nlp-chi1a} for  $\tilde{\chi}_2^{1a}$, 
\begin{align}
\tilde{\chi}_2^{1b}\left({\bf b}\right)\ &=\ 8(GM_{\sigma})^2E_{\phi}{1\over b^{\perp}}\left({15\pi \over 32}\right),  \\
\tilde{\chi}_2^{1a}\left({\bf b}\right)\ &=\ 0\ (\mathrm{4\ dimensions)}\, .
\end{align}
These, along with Eq.\ (\ref{eq:nlp-chi1a}) for $\chi_2^{1a}$ in $4-2\epsilon$ dimensions, are our basic results. The first contribution, $i\tilde{\chi}_2^{1b}$, has been known for some time, in particular, see \cite{Will}, where it appears as the first post-Newtonian correction to the calculation of the deflection of light by a spherically symmetric body in general relativity. It is also in agreement with Refs.\
\cite{Bjerrum-Bohr:2016hpa,Bjerrum-Bohr:2014zsa} and is consistent with \cite{DAppollonio:2010krb}. In the language of Feynman diagrams, one obtains the scattering of a fast light particle from a Schwarzschild metric at leading order, in accordance with the early work of \cite{Duff:1974ud}. The second correction, the result for $i\tilde{\chi}_2^{1a}$, starts at the two loop level and has appeared in Ref.\ \cite{DAppollonio:2010krb}, as noted above.  For arbitrary dimensions, both these contributions are of the same order in the ratio $\Delta/E_{\phi}$. The seagull and graviton triangle corrections are expected to exponentiate since they appear to be contributions to the gravitational Wilson line to this order.   The exponentiation properties of $\chi_2^{1a}$ are not immediately given by the reasoning here.

We may now see how the power correction $\chi_2^{1b}$  affects the amplitude in four dimensions, by transforming back to momentum space. Assuming the exponentiation of $i\tilde{\chi}_2^{1b}$, we start by rewriting Eq.\ \eqref{expon} as
\begin{align}
i\widetilde{\mathcal{M}}^{0+1}\approx 2(s- M_\sigma^2)[e^{i(\chi_0+\chi_2^b{1})}-1]\, ,
\end{align}
to find the correction to the stationary phase point, Eq.\ \eqref{saddle_b1} due to $\chi_2^{1b}$,
\begin{align}
\mathcal{M}^{0+1} \left({\bf\Delta}\right) &=
\nonumber\\
& \hspace{-15mm} 2(s- M_\sigma^2)\ \int d^2{\bf b} e^{i{\bf \Delta}\cdot {\bf b}} \ [e^{i(\chi_0+ \chi_2^{1b})}-1]\, .
\end{align}
As usual, the condition ${\partial \over \partial b}$Phase$=0$  determines the new saddle point.  In terms of $R_s=2GM_\sigma$, this condition reduces to
\begin{align}
&\Delta (b)^2\ -\ 2R_s E_\phi b\
 -\ \left (\ \frac{15\pi}{32} \right )\ 2R_s^2 E_\phi =0\, .
\end{align}
The relevant solution, at large impact parameter, is
\begin{align}
b\sim  R_s \ \left( \frac{2E_\phi}{\Delta} + \frac{15\pi}{64} \right)\, .
\label{saddlepoint}
\end{align}
The first term on the right is the leading order saddle point for the impact parameter, and the second term is the correction due to the next to leading power eikonal phase. While the leading order term shows that the impact parameter differs from the Schwarzschild radius of the heavy particle  by a large factor of $E_{\phi}/ \Delta$, no such enhancement is present for the correction.   
As discussed in Ref.\ \cite{Bjerrum-Bohr:2016hpa,Bjerrum-Bohr:2014zsa}, the same saddle point equation reproduces the deflection of light in a gravitational field.

Thus as expected, the leading order result suggests that the scattering process is dominated by large values of the impact parameter for near-forward scattering. Nevertheless,  the power correction to the leading eikonal result shifts by a finite factor times the Schwarzschild radius.  Corrections are proportional to the ratio of momentum transfer to projectile energy, and are independent of the heavy particle mass.   This suggests that for the system under study,  small angle scattering is a transition region, where corrections to eikonal propagation and gravitational self-interactions can be of comparable sizes, although at this power the former vanish in four dimensions.   In this light it may be interesting to look at yet higher powers in the our expansion about the eikonal approximation \cite{White:2011yy}.


\section{Conclusions}
\label{Disc}

In this paper we have analyzed the eikonal limit for the gravitational scattering of a high-energy or massless scalar particle by a very heavy scalar.   We have shown the self-consistency of the eikonal expansion, and have shown that the exponentiation of the leading eikonal applies as well for the first non-leading power in the energy of the light particle.  
We have calculated explicitly the next-to-eikonal contribution to this gravitational scattering amplitude. 

Working at leading power in the heavy particle mass, we expanded the light particle propagator to next-to-eikonal power and included gravitational interactions of comparable size, finding power corrections suppressed by a single power of $\Delta/E_{\phi}$ with respect to the leading eikonal term.  We have presented the non-gravitational expansion in arbitrary dimensions, where they agree with previous calculations \cite{DAppollonio:2010krb}.

Corrections are pure phases, leaving leading-power exponentiation unaffected, and are consistent with the exponentiation of the power corrections themselves.   The next-to-eikonal corrections vanish in four dimensions, while gravitational corrections cause the saddle point of the impact parameter to shift in magnitude by an amount comparable to the Schwarzschild radius of the heavy particle. Our calculations are of relevance for the small angle scattering of a light particle from a black hole.    We hope it will also set the stage for the study of further power corrections in diagrammatic or related analyses.

\acknowledgements
R.S. thanks the Rackham Graduate School for their generous support. The work of G.S. was supported by the National Science Foundation, grants PHY-0969739, 1316617, 1620628 and 1915093.  In the preparation of the revised  paper, we have benefited from discussions with Gabrielle Veneziano, Paolo Di Vecchia and John Donoghue. We would like to thank the authors of \cite{Bjerrum-Bohr:2016hpa} for sending us a copy of their paper before submitting to the arXiv. 


\bigskip

\bibliographystyle{utphys}

\end{document}